\documentclass[fleqn,usenatbib]{rasti}
\usepackage{newtxtext,newtxmath}
\usepackage[T1]{fontenc}
\DeclareRobustCommand{\VAN}[3]{#2}
\let\VANthebibliography\thebibliography
\def\thebibliography{\DeclareRobustCommand{\VAN}[3]{##3}\VANthebibliography}
\usepackage{graphicx}	
\usepackage{amsmath}	
\usepackage{subfig}
\usepackage{soul} 

\def\apgt{\ {\raise-.5ex\hbox{$\buildrel>\over\sim$}}\ }

\newcommand{\MSun}{\mbox{M$_\odot$}}

\title[RL in cluster dynamics simulations]{ReLaTS: a Reinforcement
  Learning-based method for dynamically determining the coupling
  Time Step in multi-scale simulations of self-gravitating systems.}

\author[Veronica Saz Ulibarrena et al.]{
Veronica Saz Ulibarrena,$^{1}$\thanks{E-mail: veronica.saz.ulibarrena@gmail.com}
Simon~Portegies~Zwart,$^{2}$ \\
$^{1}$Leiden Observatory, Leiden University, Einsteinweg 55, 2333 CC, Leiden, The Netherlands
}

\date{Accepted XXX. Received YYY; in original form ZZZ}

\pubyear{2025}

\begin{document}
\label{firstpage}
\pagerange{\pageref{firstpage}--\pageref{lastpage}}
\maketitle

\begin{abstract}
Astrophysical simulations frequently address multi-scale,
multi-physics problems through subsystem decomposition,
problem-tailored integration schemes, and coupling on fixed manually set timescales. Here we introduce ReLaTS, a reinforcement learning framework that dynamically selects the coupling time step to optimize the trade-off between accuracy and computational cost.
We validate ReLaTS on star clusters containing a planetary system, and test the method by varying the number of stars $N_\star$ in the cluster and the number of planets ($N_{\rm planet}$) orbiting one of them. The method finds the optimal coupling time step that balances speed and accuracy without requiring expert knowledge. In addition, the trained network operates independently of the coupled \textit{N}-body algorithms, displaying stable performance across a range of setups. We observe that the method is less reliable for cases with infinitesimal masses, as their contribution to the total energy is negligible compared to that of the massive bodies, and the network is not capable of recognizing potential errors generated while integrating them.
For long-time integration of large $N$ systems, the error accumulates.
The reinforcement learning algorithm, however, manages to keep the
energy error below a pre-set threshold. This approach substantially
reduces energy errors relative to fixed-time step baselines without
substantial additional computational overhead. Once trained, ReLaTS requires no expert tuning and generalizes across diverse astrophysical domains, enabling adaptive multi-scale simulations.
\end{abstract}



\section{Introduction}
\label{Introduction}

There is a growing demand for simulations to cover a wide range of scales and physical processes. The various scales or processes can be
numerically addressed by hybridizing the numerical solver. Each physical process, or each separable scale, is then addressed with a
different dedicated solver.  The coupling of these solvers is realized
with a bridging algorithm \citep{fujii2007bridge}. The dedicated
solvers address a limited dynamic range and a limited palette of
physical processes, but are generally optimized for performance and
accuracy. The bridging strategy, however, is much harder to
optimize. It is typically unaware of the differential equations or the
numerical strategies adopted for solving them. The danger in these
coupled simulations is that even though each dedicated solver performs
excellently, the coupling strategy fails by being insufficiently aware
of how the coupling is optimally addressed.

Tuning the bridging algorithm in such coupled problems requires
considerable expert knowledge of the underlying physics, the algorithms, and the implementation of the dedicated solvers. The coupling strategy and the associated (tuning) parameters depend
sensitively on the coupled codes, the physical processes one tries to simulate, the range of scales in the problem, and the topology of the distribution functions.  At the same time, the results of such a hybrid simulation are hard to validate in terms of scientific interpretability. One cause of this difficulty lies in the non-linearity of the results produced by dynamically coupled differential solvers. 
In addition to the non-linear response of such coupled systems, they are also intrinsically chaotic. The combination of the non-linear propagation of errors and the intrinsic chaotic behavior makes it a fantastic, challenging computational problem; it dramatically complicates one's ability to validate the results of such simulations.

Take, for example, as we do in this paper, the dynamical evolution of a planetary system in a stellar cluster.  This is a self-gravitating
system with (many) more than 2 particles, and it exhibits a wide range
of scales.  This system complies with all the above complexities by
being chaotic (the gravitational $N$-body problems for $N>2$ is chaotic, and the planetary dynamics is generally well separated from the global cluster dynamics, typically by more than 3 or 4 orders of magnitude.

We measure accuracy in the simulation by keeping track of the total energy error, see section \ref{sect:error_evaluation}.  For the direct
integration of Newton's equations of motion, the computational effort scales with $N^2$, where $N$ is the number of bodies in the system. As a consequence, large $N$ systems can become quite expensive to integrate in terms of computer time \citep{2003gmbp.book.....H,Aarseth2003}.

The smallest time scale in the system dictates the coupling time, and the computing time then scales with $N^2$. If the system exhibits a wide range of spatial or temporal scales, it becomes inefficient to integrate all particles on the same smallest scale (\cite{1975ARA&A..13....1A,2012NewA...17..711P}). Therefore, we separate this dynamic range into a small (child) system and a large
(parent) system. In a star cluster, this separation in scales is
naturally provided in the planetary system as the child, and the
cluster at large as the parent. Time scales in the child system are on
the order of years, whereas time scales in the parent system are on
the order of millions of years.  By integrating the parent system with
internally relatively large (multi-millennia) time steps, we can
integrate the child (planetary) system on a time scale of sub-years
while only resolving the interaction between parent and child on the
coupling time scale.  We call this \texttt{Bridge} (\cite{fujii2007bridge}).

Apart from enabling the strict separation of child and parent systems,
\texttt{Bridge} also allows us to choose a different integrator for each of them. Our naive implementation of \texttt{Bridge} fails when objects move from one domain to the other, for example, when a planet is ejected from the planetary system. In this case, the planet should eventually be removed from the child system to be incorporated into the parent system. This logic is not implemented in the experiment discussed in this paper, but it is supported in \texttt{Nemesis} \citep{zwart2020non}, a global structure that couples the stars and planetary system dynamics through a cascade of \texttt{Bridge} patterns.


In our simpler setup, both parent and child are coupled through \texttt{Bridge}, and the accuracy is controlled through the \texttt{Bridge} time step $\Delta t_{\rm B}$.  The problem lies in deciding the value of $\Delta t_{\rm B}$ that gives the most accurate results for the smallest computational overhead. Too large a value of $\Delta t_{\rm B}$ will make the calculation very fast, but inaccurate, whereas a too small value of $\Delta t_{\rm B}$ makes the calculation unnecessarily (or even impossibly) slow.

So far, $\Delta t_{\rm B}$ is determined for each experiment by an expert, and remains constant throughout the simulation. This works satisfactorily so long as the system's topology remains more or less unaffected by the dynamics\footnote{Here ``more or less'' indicates that the choice of $\Delta t_{\rm B}$ is not critically important, so long as approximately the right order of magnitude is adopted}, as explained in \cite{fujii2007bridge}. In practice, the system may change substantially with time, in particular if the system is integrated for longer than a relaxation time scale of either the parent or the child system. To integrate such systems, a constant value of $\Delta t_{\rm B}$ leads either to inaccurate results, or to unnecessarily long integration times \citep{veronicasazRL}.

This is not the first time that machine learning (ML) has been applied to addressing the gravitational \textit{N}-body problem. Earlier examples include attempts to solve Newton's equations of motion directly by training deep networks \citep{2020MNRAS.494.2465B, cai2021physics, greydanus2019hamiltonian, ulibarrena2024hybrid}.  Most of these methods focus on replacing the integrator with a deep neural network to reduce computing time.

In this work, we train a neural network to choose the best \texttt{Bridge} time step dynamically. Rather than being a trivial task, choosing an appropriate \texttt{Bridge} time step requires considerable expert knowledge, and it is not intuitive to adapt to the system's topology.

In ReLaTS, we make use of Reinforcement Learning (RL) techniques to determine the optimal cross-integration \texttt{Bridge} time step ($\Delta t_{\rm B}$) at run time. The network dynamically balances computational performance with accuracy to find the largest possible value of $\Delta t_{\rm B}$ that provides a sufficiently accurate solution. Our RL strategy typically leads to faster and more accurate results than calculations with fixed $\Delta t_B$. We tested the method on a range of problems with various numbers of stars and planets, and with different integrators for the parent and child systems.  We demonstrate that integrating the parent system using the hierarchical tree algorithm
(\cite{barnes1986hierarchical}) gives excellent results.

\section{Methodology}
\label{sec:methodology}

We couple two different parts of a multi-scale system while integrating a reinforcement learning method to determine the time-scale on which both separately-solved systems are mutually integrated.

\subsection{System setup}
\label{subsec:bridge}

We focus on a stellar cluster in which one of the stars is orbited by a planetary system.  Integrating the cluster as well as the planetary
system becomes quadratically more expensive as the number of bodies
increases. Additionally, the computational time increases linearly with the time step size in the system. Since the cluster is more than $\sim 10^3$ larger than the planetary system, so are the time step sizes that must be used for the integration of each of them. Therefore, the computer time is completely dominated by the integration time step required for the tightest planet orbiting its host. The typical time step size for a planet orbiting a star is a few days, whereas integrating a star crossing the cluster easily relaxes to time step sizes of millennia; a temporal scale difference of $10^5$.  We therefore investigate here the strict separation of the cluster dynamics from the planetary dynamics, for which we adopt \texttt{Bridge}.

Both parent (cluster of stars) and child (star orbited by multiple
planets) are integrated with a separate numerical solver.  The most
natural choice for the parent system is a high-order direct \textit{N}-body
strategy, whereas for the planetary system, we adopt a symplectic
method.  The parent and child systems are subsequently integrated
through \texttt{bridge} \citep{fujii2007bridge, zwart2013multi}. We adopt the implementation presented in AMUSE
\citep{2018amuse,2009NewA...14..369P,zwart2013multi,2026araa.book.....P}.
  
\texttt{Bridge} allows us to combine two (or more) separately integrated sub-systems into a single self-consistent solution. Every \texttt{Bridge} time step $\Delta t_B$, the acceleration from the child onto the parent is calculated, and vice versa. These accelerations are then used to correct the velocities of all particles in the system. Afterwards, the separate parts are evolved individually. This procedure is repeated until the final simulation time is reached.  

One drawback of \texttt{Bridge} is the coupling time scale $\Delta t_B$, which is manually set. This coupling time should be selected to balance accuracy and computation time, and it should ideally vary along the simulation. However, determining an optimal value for $\Delta t_B$ requires expert knowledge. Another drawback of \texttt{Bridge} emerges when a particle of one of the separate systems enters the domain of the other. This can happen, for example, when a planet is ejected from its host star to become a rogue planet in the star cluster, or when another star enters the planetary system. In those cases, the rogue object should migrate from one \texttt{Bridge} domain (the child in the case of the rogue planet) to the other (in this example to the parent system).

\texttt{bridge} explicitly assumes that the various systems remain strictly separated. In practice, however, this is not always the case. More complex coupling algorithms like \texttt{Nemesis} \citep{zwart2020non} allow for objects to move from one domain to the other, although the underlying algorithm is still fundamentally based on \texttt{Bridge}. Although we do not make use of this method, the reinforcement learning strategy to determine $\Delta t_B$, could be easily adapted for its use with \texttt{Nemesis} (we discuss this further in section \ref{Sec:Discussion}).

\begin{figure}
	\centering
	\includegraphics[scale=0.7]{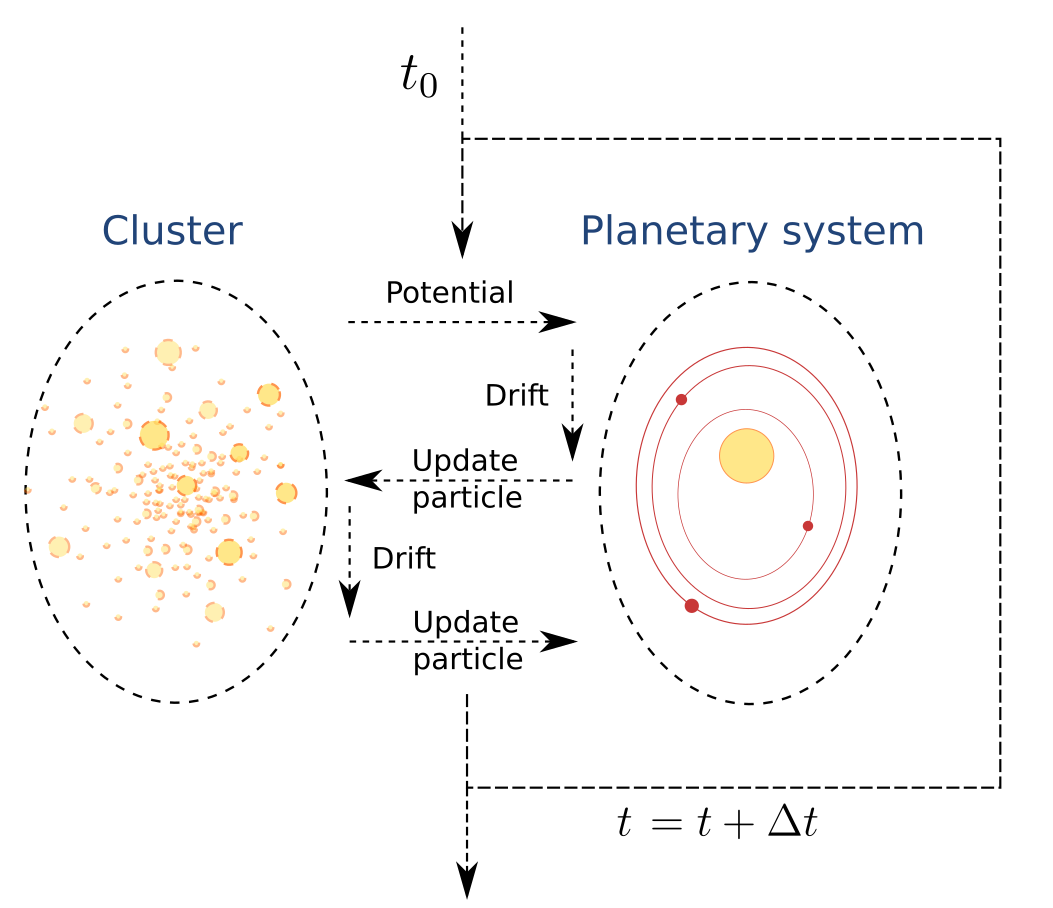}
	\caption{Schematic of the \texttt{iBridge} method as developed for this application. The system is divided into two parts: a star cluster and a planetary system, with one star being common for both parts. The acceleration caused by the cluster on the planetary system is calculated and used to update the velocities of the bodies in the planetary system. Then the planets and central star are evolved, and the state of the central star in the cluster is updated with that of the planetary system. Afterwards, the cluster is evolved and the state of the center of mass of the planetary system is updated using the latest cluster information. This process is repeated at every step. }
	\label{fig:modifiedbridge_schem}
\end{figure}

Because of this strict separation of the various systems involved in \texttt{Bridge}, we encounter one main limitation: the star that contains a planetary system should be integrated both in the parent system (star cluster) and the child system (planetary system). To address this, we introduce {\em inclusive Bridge} (or \texttt{iBridge}), a classic \texttt{Bridge} method in which one particle is common in both the parent and child systems. In Figure \ref{fig:modifiedbridge_schem} we illustrate \texttt{iBridge} for a
cluster of stars in which one star is orbited by multiple planets.
Here, the parent is the cluster of stars (left in Figure
\ref{fig:modifiedbridge_schem}), and the child is the planetary system
(right).

\texttt{iBridge} starts by calculating the potential of the cluster at
the location of the planetary system, and uses this to update the
velocities of the planets and the central star. We then evolve
the planetary system using the child integrator for a time
$\Delta t_B$. Afterwards, we update the state of the common particle in the cluster. When integrating the parent system, we ignore the effect of the planets, and evolve (drift) the start cluster for a time $\Delta t_B$. Finally, we use the latest state of the star cluster to update the particles in the planetary system by drifting its center of mass. This process is repeated until the next diagnostic output time, for the end of the run.

\begin{figure}
	\centering
	\includegraphics[scale=0.35]{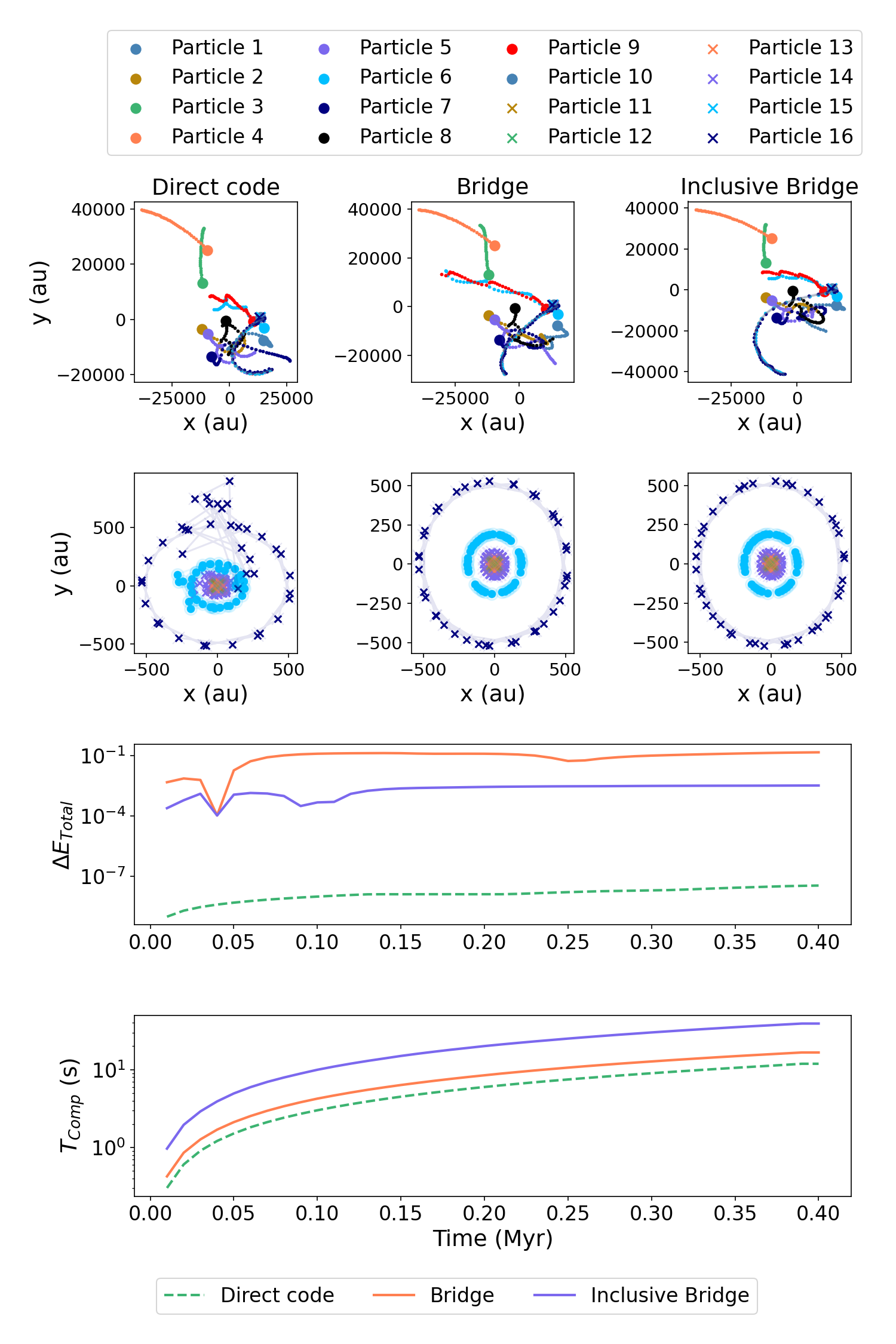}
	\caption{Comparison of our \texttt{iBridge} against the \texttt{Bridge} method as implemented in AMUSE and direct integration.}
	\label{fig:modifiedbridge_comparison}
\end{figure}

In Figure \ref{fig:modifiedbridge_comparison}, we compare
\texttt{iBridge} with \texttt{Bridge} as implemented in AMUSE, and with the direct integration of all particles in a single numerical solver. \texttt{iBridge} outperforms the regular \texttt{Bridge} in accuracy by several orders of magnitude, but it is slower by about a factor of two. Both implementations of \texttt{Bridge} generate energy errors many orders of magnitude larger than when the entire system would have been integrated directly. This is not surprising, as the direct method is optimized specifically for addressing the adopted problem with dynamically adaptable time steps. Also speed benefits of \texttt{Bridge} are negligible compared to direct integration, which is a consequence of the small number of particles in the problem being insufficient to hide the overhead of \texttt{Bridge}: we present a more detailed study of the performance of \texttt{iBridge} in Figure \ref{Fig:direct} of section
\ref{Sec:Discussion}.

For the training of the RL algorithm, we limit ourselves to \textit{N}
between 5 and 20, and use larger values of N for validation and
demonstration of the transfer knowledge capabilities of our method.

\subsection{Experimental setup}
\label{subsec:scientificsetup}

We train ReLaTS by integrating a cluster with 5 to 20 stars, including a planetary system orbiting one of the stars. Note that the entire system is solved using Newtonian gravity; we ignore relativistic
effects, tides, and radiation processes.

The masses of the stars are randomly selected from the Salpeter
power-law \citep{salpeter1955luminosity} between 1\,\MSun\, and
100\,\MSun. The stellar positions are taken from a virialized fractal
cluster model introduced by \citep{goodwin2004dynamical} with fractal
dimension of $F_d = 1.6$, and a virial radius of 0.1\,pc.  A system of
planets is introduced around a randomly selected star. The planets
follow circular orbits in a random plane, and their masses and
semi-major axis are determined from the oligarchical planetary growth
model \citep{0004-637X-807-2-157, 0004-637X-581-1-666} assuming a disk
with a mass of 0.02\,\MSun\, between 10\,au and 100\,au.  These
initial conditions are summarized in
table~\ref{table:initialconditions}.

\begin{table}
	\vskip 0.15in
	\begin{center}
		\footnotesize
		\begin{sc}
			\caption{Initial conditions and integration settings of a cluster with one planetary system orbiting one of the stars. }
			\label{table:initialconditions}
			\hspace{-20pt}
			\begin{tabular}{lc}
				\textbf{Star cluster}&\\
				\hline
				Number of stars & [5 - 20]\\
				Mass range of the stars & [1, 100] \MSun \\
				Radius of the cluster & 0.1 pc\\
				Virial ratio & 0.5 \\
				Fractal dimension & 1.6\\
				\hline 
				&\\
				\textbf{Planetary system}&\\
				\hline
				Inner disk radius & 10 au\\
				Outer disk radius & 100 au\\
				Disk mass         & $0.02$\, \MSun \\
				\hline
				&\\
				\textbf{Integration}&\\
				\hline
				Cluster code & Ph4\\
				Cluster code $\eta_C$ & $10^{-2}$\\
				Planetary system code & Huayno\\
				Planetary system code $\eta_P$ & $10^{-2}$\\
				Training step size & $10^{-2}$ Myr\\
				\hline
			\end{tabular}
		\end{sc}	
	\end{center}
	\vskip -0.1in
\end{table}

We integrate the parent system with the 4th order Hermite
predictor-corrector integrator implemented in the code {\tt ph4}
\citep{2022A&A...659A..86P}. For the child (planetary system) we adopt
the symplectic connected components method {\tt Huayno}
\citep{janes2014connected}.  {\tt Huayno} is derived from 2nd order
Hamiltonian splitting for \textit{N}-body dynamics, which makes it
well suited for the integration of planetary systems. Each of these
codes is incorporated in {\tt AMUSE} \citep{2026araa.book.....P}.  Later in
section\,\ref{sec:tree}, we will demonstrate that {\tt ReLaTS} produces robust results, generalizes, and works independently of the selected integrator for the parent or child systems.  Each code is initialized with a time-step parameter ($\eta$) as indicated in table
\ref{table:initialconditions}, which scales the internally-calculated
time-step sizes. The state of the system is saved, and the \texttt{Bridge} time step re-evaluated with a frequency determined by the check step size.

There are four main time steps involved in this setup:
\begin{itemize}
	\item \textbf{Time-step size of the cluster integration}: this is
      the time step used for the integration of the star cluster. When
      using Ph4, this time step is calculated internally and
      multiplied by a scaling parameter ($\eta_C$).
	\item \textbf{Time-step size of the planetary system integration}:
      this is the time step used for the integration of the planetary
      system. When using Huayno, this time step is calculated
      internally and multiplied by a scaling parameter ($\eta_P$)
      which might be different from $\eta_C$ used for the cluster.
	\item \textbf{Bridge time step} $\Delta t_B$: this is the time
      scale on which the parent and child exchange information. This
      means how often the method calculates the acceleration caused
      on one system by the other and updates the common particle. This
      time-step size has to be selected at the start of the
      simulation. There is currently no implementation that
      automatically selects an optimum value for $\Delta t_B$; this
      selection is done based on expert knowledge. This is the time
      step that we will determine using RL.
	\item \textbf{Training step size}: time scale used to save the state
      of the system and apply the reinforcement learning method. This
      means after how much time the choice of the \texttt{Bridge} time step is
      re-evaluated.
\end{itemize}

\begin{figure*}
	\centering
	\includegraphics[width = 2\columnwidth]{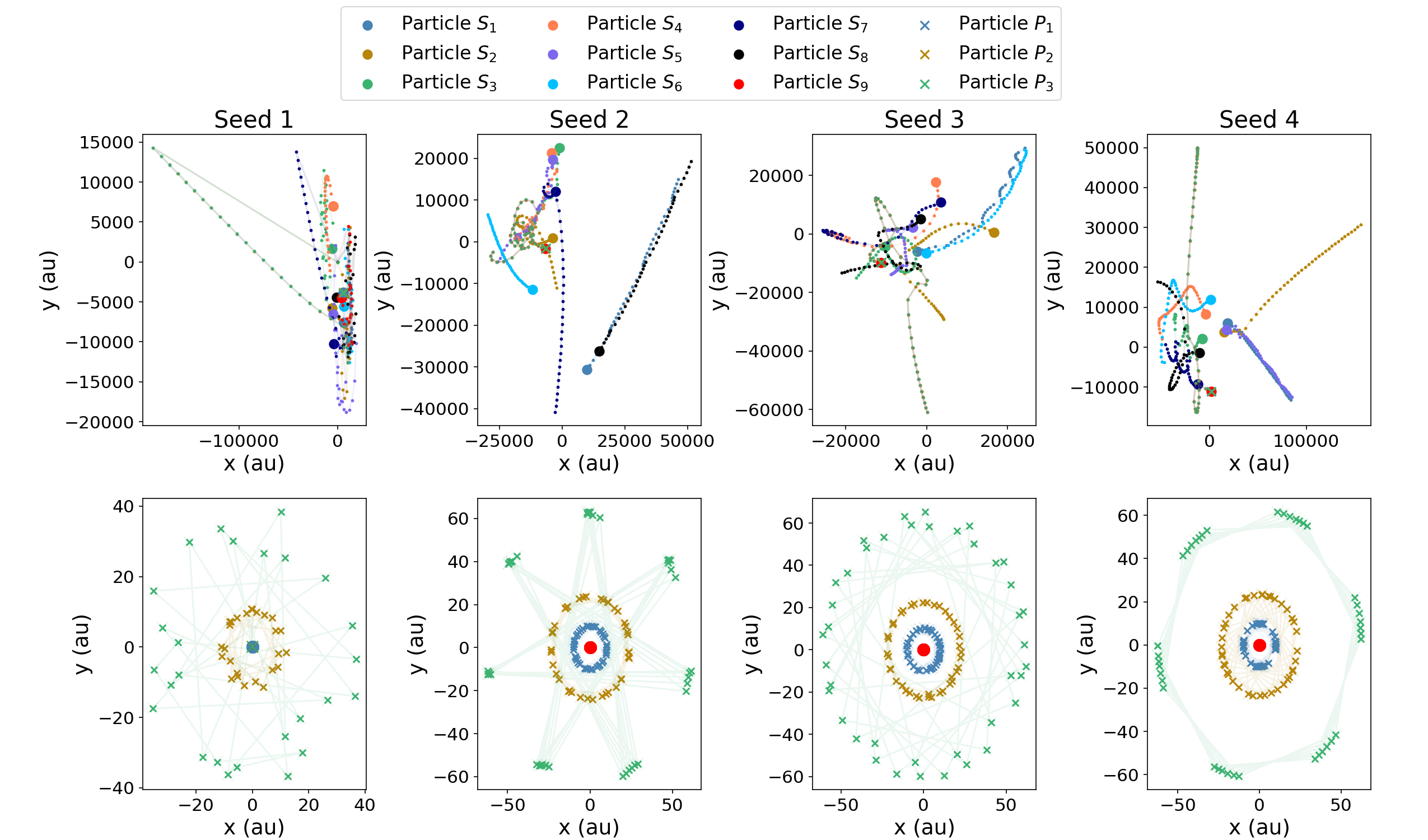}
	\caption{Initializations for Seeds 1 to 4 run for 40 steps (0.4
      Myr) with a \texttt{Bridge} time-step of $5 \times10^{-5}$
      Myr. The setup is formed by 9 stars and 3 planets. }
	\label{fig:initialconditions}
\end{figure*}

In Figure \ref{fig:initialconditions} we present four examples of the
integration of the initial conditions from Table \ref{table:initialconditions} with 9 stars for four different values
of the random number seeds; we call them seed 1 to 4. The top row
shows the evolution of the positions of the stars in the cluster. The
bottom shows the evolution of the planets. The stars are represented
by bullet points and the planets by an ``x''. This system experiences large differences in its evolution depending on the initial realization (the seed). As a consequence, the optimal \texttt{Bridge} time-step size differs depending on the initial realization. In addition, since the local conditions change rapidly with time, the optimal \texttt{Bridge} time-step varies during the simulation.

\subsection{Validation of the simulations through the energy error}\label{sect:error_evaluation}

Numerical errors in simulations of self-gravitating systems originate
from round-off at the least significant digit, and discretization
errors due to the finite time step adopted in the numerical
scheme. Such infinitesimal errors (round-off typically occurs around
the 15th decimal place, and time discretization errors around the 10th
decimal place) grow exponentially due to the chaotic characteristics
of the systems. The numerical errors in a chaotic system inevitably render the system unique after an e-folding time scale (or Lyapunov time). As a consequence, our  experiments cannot be validated by performing a simulation to convergence \citep{2015ComAC...2....2B}, \footnote{A converged solution is a
solution for which the realization of the system at a specific moment
in time is independent of the numerical integrator or round off.}
because both solutions have diverged and can only be compared
statistically.

Since the phase space (in position and velocity) cannot be used for
validation purposes, the only remaining diagnostics are the conserved
quantities; energy, linear momentum, and angular momentum. Since many
numerical methods are designed to conserve linear and angular momentum, we opt for the conservation of energy as the remaining tool for validation purposes. Note that at the same total energy, the
self-gravitating \textit{N}-body problem still has infinitely many
configurations with very different phase‑space trajectories and
spatial structures. Energy conservation alone does not guarantee the
right solution, but at least it gives a solution that theoretically
can be reached by the initial conditions in an ergodic problem.

High-order integration strategies, such as the one adopted here for the parent system, typically conserve energy to $10^{-9}$ (or better) per step. The energy error, however, tends to drift with time, causing long time-scale integrations to progressively move away from the actual system.

A symplectic and time-symmetric integrator, such as the 2nd-order
method adopted here for the child system, keeps energy nearly constant
over long times but still exhibits errors in orbital phase (a quantity
not naturally conserved in physics). The relative energy error then
remains the natural choice for the validation of the performance of a
numerical integrator for Newton's \textit{N}-body problem (cf. the discussion in Section \ref{Sec:Discussion}).

In practice, the total energy of the system is not perfectly conserved
during the simulation; the use of numerical integrators leads to
energy errors. The total energy is the sum of the kinetic and
potential energy of the system. We define the energy error as the
relative difference of the energy at time step $i$ compared to the
initial realization:
\begin{equation}
\Delta E_i = \dfrac{(E_{k,i} + E_{p,i}) -(E_{k,0} + E_{p,0})}{E_{k,0} + E_{p,0}} = \dfrac{E_i-E_0}{E_0}.
\end{equation} 

A large value of $\Delta E_i \apgt 10^{-4}$ is an indication of an
unphysical solution, while infinitesimal values of $\Delta E_i$
show appropriate behavior, and we adopt the energy error as an
unbiased measurement of the simulation's accuracy to validate the
results and training the network. 

\subsection{Reinforcement Learning}
\label{subsec:RL}

We train the reinforcement learning in ReLaTS to determine the optimal
\texttt{Bridge} time-step $\Delta t_B$. We choose Q-learning as our RL
algorithm to maximize a reward value \textbf{\textit{R}}
(\cite{sutton2018reinforcement}).  By supplying the algorithm with
information from different astronomical simulations, it learns to take
actions that maximize the reward. We adopt Q-learning for its relative
simplicity and demonstrated efficiency. One of our objectives is to
understand how RL methods can be combined with complex astronomical
problems. Therefore, we choose the simplest implementation and leave
further studies and comparisons to more advanced RL methods for future
work.

We adopt Deep Q-networks (DQN), an extension of Q-learning dedicated
for continuous states, as is the case for the state of the system we
study.  Deep Q-networks combines reinforcement learning with deep
neural networks \citep{mnih2015human}; they are used to approximate
the optimal value of the Q-function
\begin{equation}
Q(S, A) = \max_\pi \mathbb{E} R_t + \gamma R_{t+1} + \gamma 2 R_{t+2} + \ldots,
\end{equation}
which is the maximum sum of the rewards multiplied by the discount
value $\gamma$ at each time step $t$. This maximum is chosen following
a behavior policy $\pi = P (A|S)$ after making an observation $S$ with
an action taken $A$. To balance the trade-off between exploration and
exploitation, the algorithm includes a stochastic exploration strategy
called $\epsilon$-greedy, by which a random action is chosen with
probability $p$ instead of the one selected by the algorithm
\citep{Viquerat2022DRLfluid}. This value $\epsilon$ is reduced during
the training to favor exploitation over exploration. To avoid the
inherent instabilities of RL, the method uses experience replay, which
stores the data in a training database and randomizes the chosen
training sample to eliminate correlations in the observation sequence
\citep{mnih2015human}.  We avoid instabilities and variability during
training by employing two different networks
(\cite{yu2018historical}); the DQN, where the weights are updated at
each training step, and the Target net, which is only updated with the
weights of the Q-net after a given number of steps.
	
One limitation of our adopted method is its discrete action
space. Ideally, we'de employ a continuous action space, which can, in
principle, be achieved using the Soft-Actor Critic \citep[or
  SAC,][]{haarnoja2018soft}) strategy. However, the simplification of
the action space into a discrete one can be beneficial for the purpose
of interpretability and simplicity. This choice most likely results in
a decrease in performance as the algorithm does not have information
about the relations between actions. However, this can also be used as
an advantage to make our method more generalizable to different
astronomical scenarios. In principle, any two codes can be bridged
with ReLaTS, so long as the numerical error in the energy provide an
adequate diagnostics for evaluating the \texttt{Bridge} time step
$\Delta t_B$.

Since the trained algorithm is not linked to the value of the action
space, but to the integers corresponding to the action numbers, this
trained algorithm can still be used for different simulation settings
(see Subsection \ref{subsec:integration}).

We base the algorithm on the work by \cite{mnih2013playing} and the Pytorch tutorials (\cite{code_RL_tutorial}). More complex methods could be used to achieve better performances. However, as the goal of this study is to gain an intuition of how RL can be used for this astronomical problem, we limit the study to the use of DQN and leave the comparison with other algorithms for future work.

\begin{figure}
	\centering
	\includegraphics[width =\columnwidth]{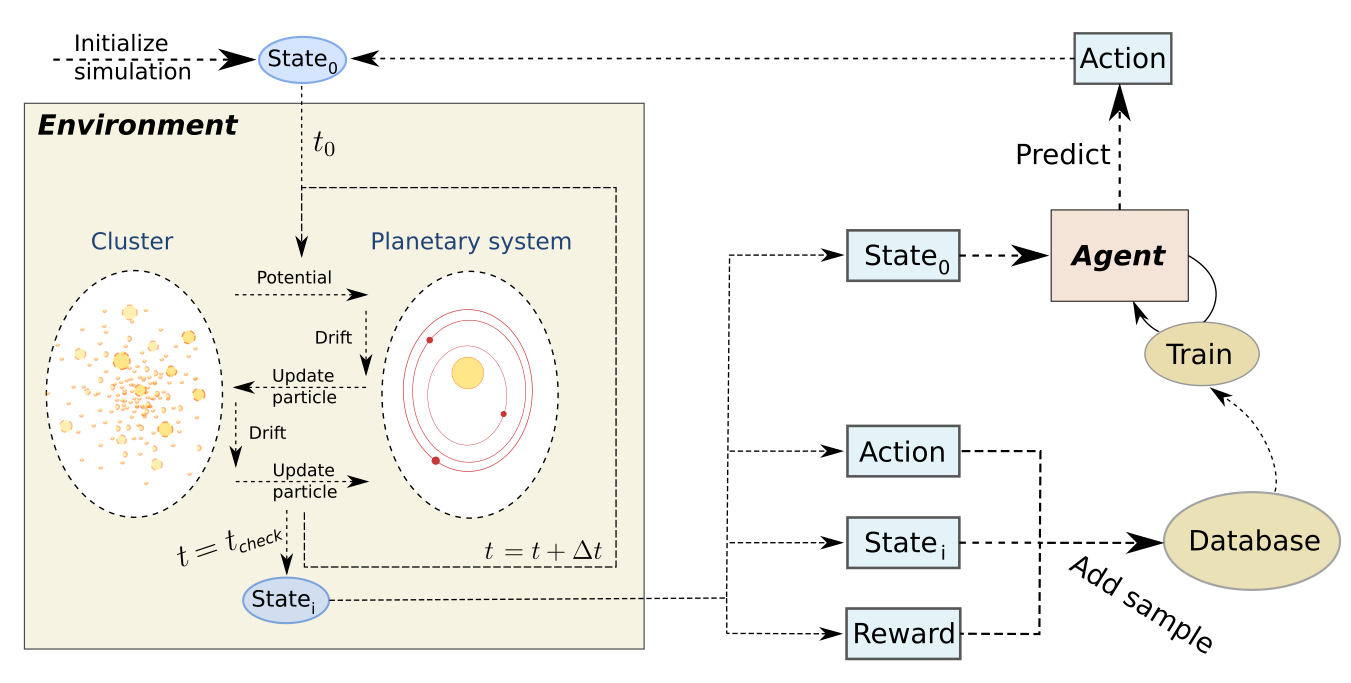}
	\caption{Schematic of the interaction between the Environment and the Agent.}
	\label{fig:clusterschem}
\end{figure}

There are different elements interacting in the DQN method, as seen in Figure \ref{fig:clusterschem}:
\begin{itemize}
	\item \textbf{Environment}: the environment is composed of the
      astronomy simulations. The data obtained from them is used to
      create a dataset of the states, rewards, and actions. The
      composition of the environment is as explained in Subsection
      \ref{subsec:scientificsetup} and the astronomical simulations
      are initialized using random seeds during the training.
	
	\item \textbf{Agent}: The agent is the reinforcement learning
      algorithm trained to select actions. It is composed of two
      neural networks, namely the Q-net and the Target net. The
      weights of the Q-net are updated at each training step, whereas
      the Target net is updated only after a fixed number of
      steps. Both networks receive as input the state
      (\textbf{\textit{S}}) of the system generated by the environment
      and produce the corresponding Q-values associated with each
      possible action.

      To balance the trade-off between exploration and exploitation,
      the training process employs an $\epsilon$-greedy strategy,
      whereby a random action is selected with probability $p$, and
      the action with the highest Q-value is selected otherwise
      \citep{Viquerat2022DRLfluid}. Thus, during training, action
      selection is stochastic to encourage exploration, while during
      inference (or pure exploitation), the action with the largest
      Q-value is selected deterministically for the next step of the
      simulation.
      The reward function is evaluated using values from the
      environment and is used to define the loss function for training
      the network.
      
	To set up the agent, we have to select the hyperparameters of the
    neural networks as well as other training parameters. The adopted
    specific values as discussed subsection \ref{subsec:Training}.
	
	\item \textbf{State}: the state is the representation of the
      environment that is used as an input to the neural networks in
      the agent. It must be formed by values that are representative
      of the physical state of the environment at a given time.
	
	In \cite{veronicasazRL}, as in many studies dealing with neural
    networks in the gravitational \textit{N-}body problem, the state
    is chosen to be the Cartesian coordinates representing the
    positions and velocities of each particle of the system. However,
    this leads to a fundamental problem: the input size is dependent
    on \textit{N}, leading to limited extrapolation
    capabilities. Although Graph Neural Networks or Autoencoders could
    provide an interesting alternative for reducing the variable state
    space into fixed-size input. We leave those implementations for
    future work, and find a simple solution for this problem. To do
    so,	we define the state (\textbf{\textit{S}}) as
	\begin{equation}
	\textbf{S} = \left[\sum_i^{N-1} V_{n_i\rightarrow n_{c}}
	\quad, \quad - \text{log}_{10}(\Delta E) \right].
	\end{equation}
	Here $\sum_i^{N-1} V_{n_i\rightarrow n_{c}}$ is the gravitational
    potential of every star in the cluster ($n_i$) at the position of
    the common star ($n_c$). The second term is the current energy
    error of the simulation; it is included to account for the
    possibility of a degenerate solution once the error exceeds some
    minimal threshold. Once this happens, it is improbable to be reduced to an acceptable. It is therefore important to
    consider this in ReLaTS in order to avoid incurring unnecessary
    computational costs.
	
	\item \textbf{Actions}: the actions (\textbf{\textit{A}}) are the
      possible values of a decision variable. The reinforcement
      learning algorithm is trained to select between these values to
      optimize a reward function.  The actions are taken from a
      finite-size array which contains the value of the control
      variable associated with each action. Our decision variable is
      the \texttt{Bridge} time-step size. At each step, an action is chosen to determine the value of $\Delta t_B$ to be taken for the next steps of the simulation. The number of actions, as well as the allowed range of values are discussed in subsection
      \ref{subsec:Training}.
	\item \textbf{Reward function}: the reward is the function to be
      optimized by reinforcement learning. For the study at hand, we
      optimize for both accuracy and computing time simultaneously. For this purpose, we design a function that balances both the energy error and the computation time. We write this function as in \cite{veronicasazRL}
	\begin{equation}
	\textbf{	R} = -W_1  \dfrac{\text{log}_{10}\left(\vert \Delta E \vert /10^{-10}\right)}
	{\vert \text{log}_{10}(\vert \Delta E\vert)\vert^3}
	+\ W_2 \dfrac{1}{\text{log}_{10}(A)},
	\label{eq:reward}
	\end{equation}
    which was specifically designed for the needs of this application.
    
	The first term in Equation\,\ref{eq:reward} corresponds to the
    energy error at a given step, normalized by $10^{-10}$ and divided
    by the cube of the energy error. This term represents a decreasing
    slope with the value approaching zero when the energy error
    approaches $\Delta E = 10^{-10}$. The logarithm is used to
    linearize the range of values in this term. We design this term
    to guarantee that the reward obtained by achieving satisfactory
    energy errors has a smaller slope than for larger errors.
	
	The second term corresponds to the computation time represented by
    the inverse of the time-step. $W_{1,2}$ are the weights used to
    balance these two terms. They are a design choice and the values
    used can be found in Subsection \ref{subsec:Training}.  This
    reward function is specifically designed for the problem of the
    simulation of a number of bodies interacting via their
    gravitational forces.  Figure \ref{fig:reward} shows the reward
    for different values of the energy error and computation time. On
    the top panel, we can see the shape of the logarithmic curve that
    penalizes large energy errors (inaccurate results), while giving a
    higher reward value to those cases obtained in lower computation
    times (shown in blue).
	
	\begin{figure}
		\centering
		\includegraphics[width = 1\columnwidth]{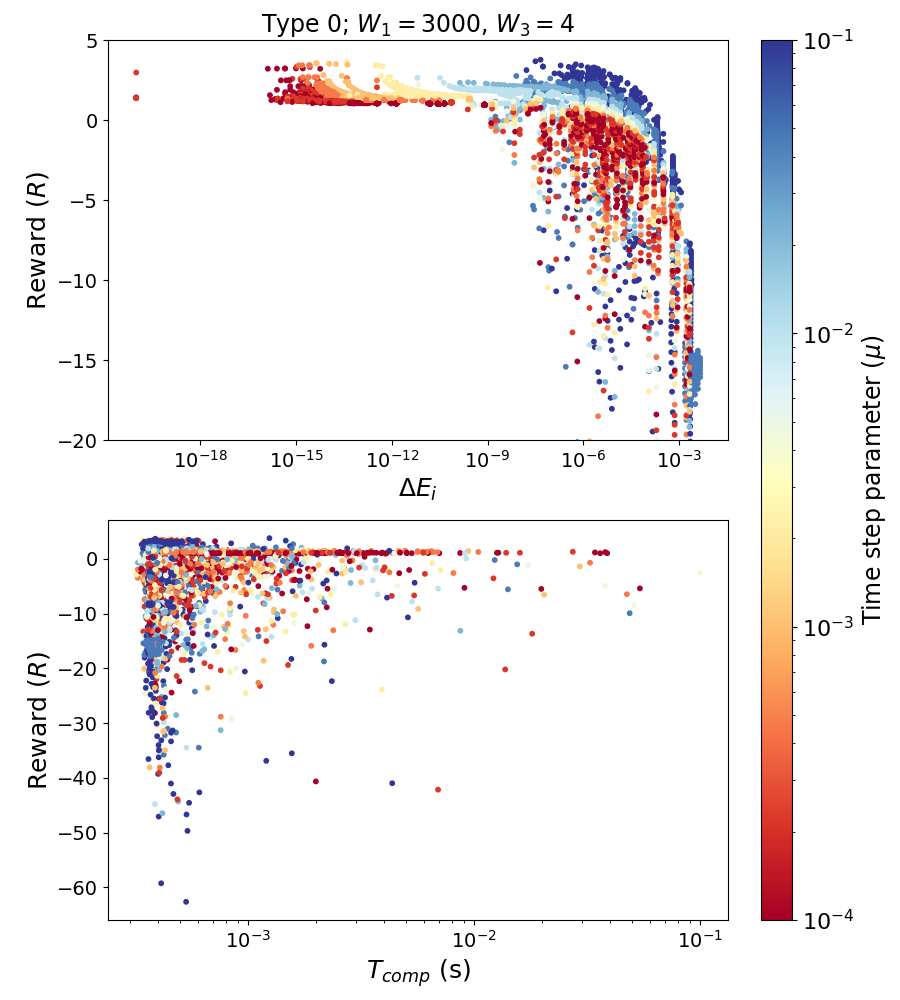}
		\caption{Reward value as a function of the energy error (top panel) and the computation time (bottom panel).}
		\label{fig:reward}
	\end{figure}
\end{itemize}


\section{Results}
\label{sec:results}
We show the results obtained from training the RL algorithm and its application to different cases of the start cluster simulation. 

\subsection{Validation of the results}
\label{subsec:convergence}

In the absence of an analytic solution to the \textit{N}-body problem, and taking into account its chaotic behavior, finding a baseline to which to compare the results is challenging. 

We therefore perform a convergence study in $\Delta t_B$ to better understand the relation between energy error and computing time, which we require to decide the range of actions (\textbf{\textit{A}}) used in ReLaTS. For this, we simulate the systems with seeds 1 to 4 (see figure \ref{fig:initialconditions}) until a time of 0.4\,Myr using different values of $\Delta t_B$.  We perform a convergence study to find the value of $\Delta t_B$ for which the energy error does not improve even when we further reduce the time step. In this limit, the computing time just increases without improving the results. This definition of convergence is different from that used for example in \cite{2015ComAC...2....2B} \footnote{From Boekholt et al.: ``A converged solution is a solution for which the first specified number of decimal places of every phase-space coordinate in our final configuration in the \textit{N}-body experiment becomes independent of the length of the mantissa and the Bulirsch-Stoer tolerance''.}.  There, a converged solution is achieved with arbitrary precision codes such as \texttt{Brutus}.

\begin{figure} 
	\centering
	\includegraphics[width=\columnwidth]{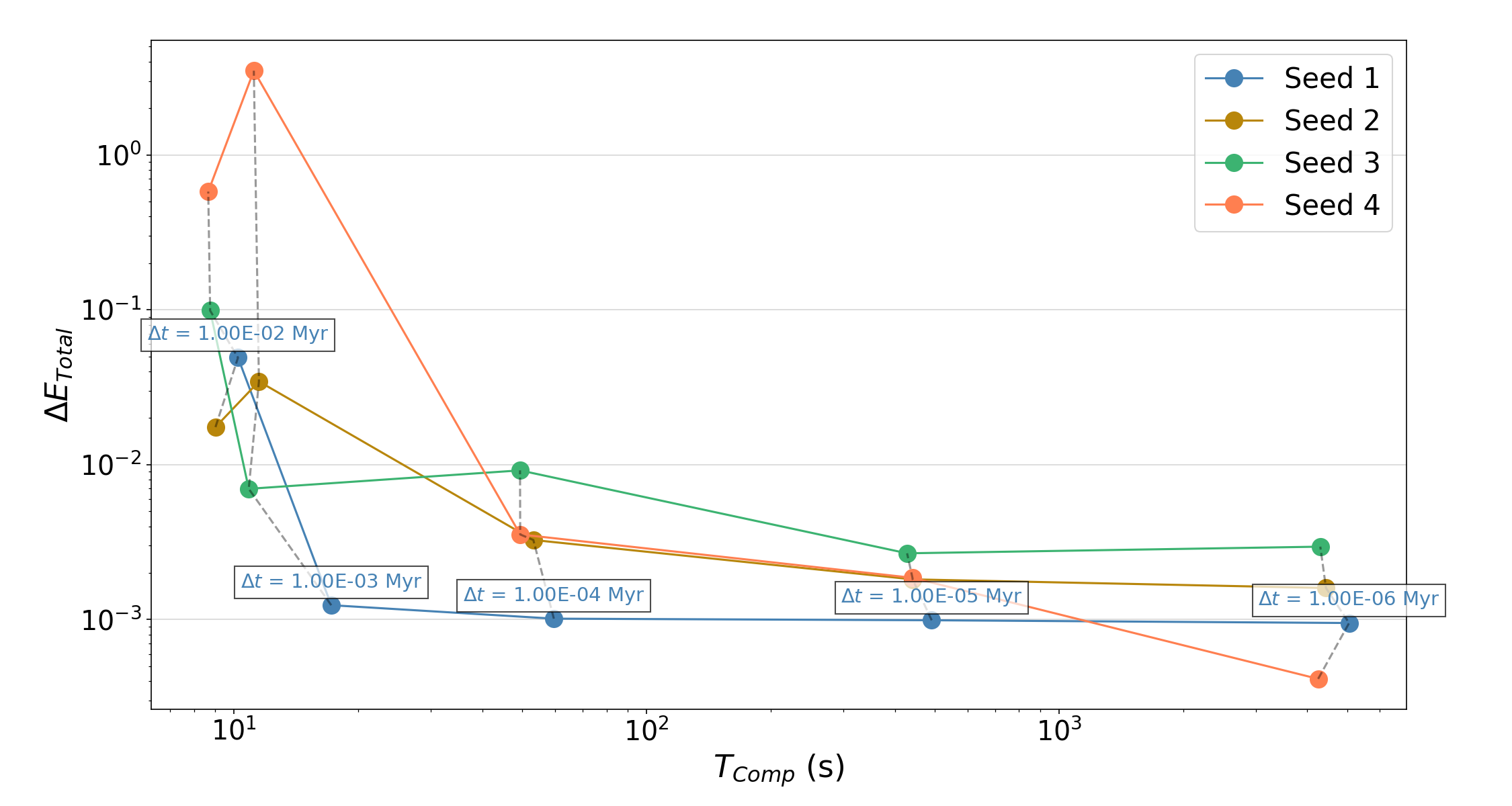}
	\caption{Energy error and computation time for the simulation of
      initializations with seeds 1 to 4 run for 40 steps (0.4 Myr) as
      a function of the constant \texttt{iBridge} time-step. The
      time-step are indicated in the figure. }
	\label{fig:convergence}
\end{figure}

We show the results in Figure \ref{fig:convergence}. The value of
$\Delta t_B$ for which the simulation converges depends on the initial
realization. Some cases such as the one with seed 4 result in the
minimum value of $\Delta t_B$ not being small enough to find
convergence in the energy error. However, for a case such as the one
with seed 1, convergence is reached for relatively large values of
$\Delta t_B$.

Although it is difficult to identify clear limits on the range of values that can be used for a specific simulation, we can observe that in most cases convergence is reached for time step sizes between $10^{-4}$ and $10^{-5}$ Myr. We therefore choose an intermediate value of $5\times 10^{-5}$ as the lowest limit for the RL actions. We see that depending on the simulation a time step size of $10^{-2}$ Myr may yield large energy errors. We choose this value as the upper limit for the RL actions.

\subsection{Training results}
\label{subsec:Training}

We describe the experimental setup in section
\ref{subsec:scientificsetup}, and discuss the RL method in section
\ref{subsec:RL}. With that information and the settings in Table
\ref{table:trainingparams}, we obtain the trained models. We set the
maximum number of integration steps to 40 which, with a check step
size of $10^{-2}$ Myr, corresponds to a final time of 0.4 Myr. The
number of planets depends on the mass of the central star, and
therefore depends on the other initial conditions. In table
\ref{table:trainingparams} we show the network architecture.

Due to the chaotic nature of the problem, evaluating the model on a
random selection of validation cases would lead to unreliable metrics
(see \cite{veronicasazRL}). In order to create a robust indication of
the performance at different episodes of the training, we test the
model on a fixed set of initial realizations. By doing so, we can
directly compare the rewards obtained on those and gain intuition
about the evolution of the training. We use 3 test cases. Ideally, this number should be increased for a better indication of the performance of the model at each episode, but due to computational
limitations, we choose to keep this number small. After training, the model is applied to a number of experiments.

The actions, \textbf{\textit{A}}, is a discrete array of length 10
with values of $\Delta t_B$ that range from $5\times 10^{-5}$ to
$10^{-2}$ Myr. Similarly to the implementation in Hermite and
Huayno, we define a time-step parameter $\eta_B$ for the
\texttt{iBridge} that multiplies the value in the actions. This value
is set by default to 1. Finally, the weights for the reward function
are shown in Table \ref{table:trainingparams} and selected such that the first term in Equation \ref{eq:reward} is 50 times larger than the
second term.

\begin{table}
	
	\vskip 0.15in
	\begin{center}
		\footnotesize
		\begin{sc}
			\caption{Training and simulation parameters.}							
			\label{table:trainingparams}
			\begin{tabular}{lc }
				\hline
				Number of planets & variable\\
				Max steps per episode & 40 \\
				$\Delta E$ tolerance & $1\times 10^{0}$\\
				\hline
				Hidden layers & 5\\
				Neurons per layer & 200\\
				Batch size & 125 \\
				Test data size & 3\\
				\hline
				Number of actions & 10\\
				Range of actions & [$5\times 10^{-5}$, $10^{-2}$] Myr\\
				$W_{1,2}$ & [50, 1]\\
				\texttt{iBridge} $\eta_B$ & 1.0 \\
				\hline
			\end{tabular}
		\end{sc}
	\end{center}
	\vskip -0.1in
\end{table}

We train the RL method in ReLaTS in several phases. First we perform a computationally inexpensive global search (high learning rate), and  reduce the learning rate in the next phase. To avoid increasing the number of tunable parameters, we keep the learning rate constant during training, rather than reducing it automatically.  Also, this step-by-step strategy allows us to perform a fast pre-training with simulations with a small number of stars (5 stars), and only then perform a local training (lower learning rate) for the more expensive simulations (larger number of stars).

\begin{figure}
	\begin{minipage}{\linewidth}
		\begin{center}
			\centerline{\includegraphics[width=\textwidth]{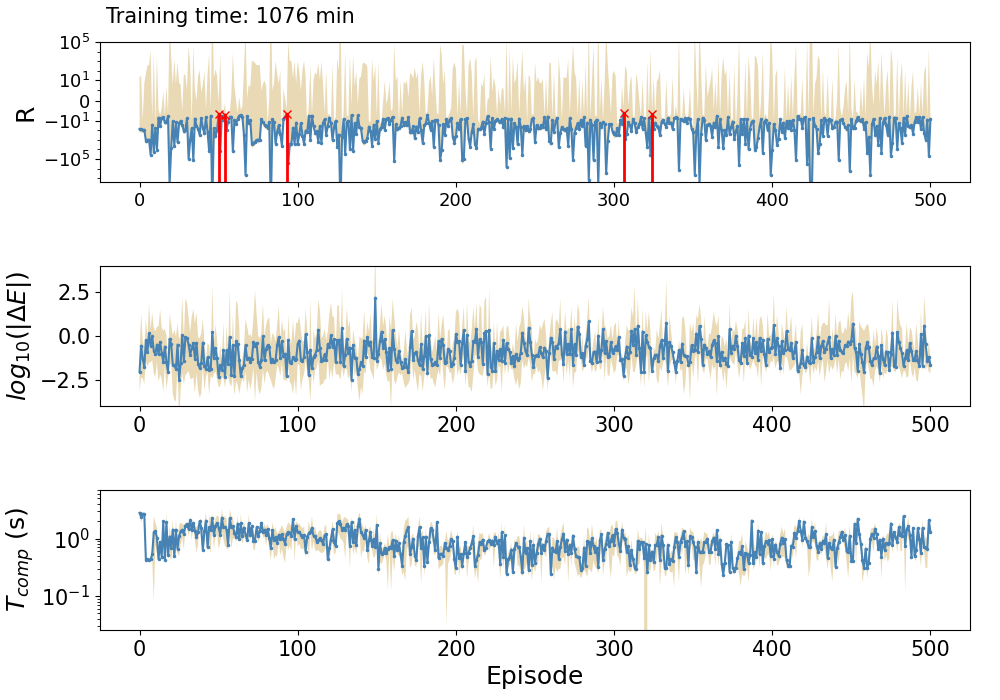}}
		\end{center}
	\end{minipage}
	\begin{minipage}{0.3\linewidth}
	\vspace{-20pt} 
	$$$$\hspace{20pt}
	\begin{tabular}{lc lc}
		\hline
		\textbf{Global search}&&&\\
		\hline
		Number of stars & 5& \hspace{20pt}
		Max episodes & 500 \\
		Learning rate & $1\times 10^{-3}$& \hspace{20pt}
		Model chosen & 50\\
		\hline
	\end{tabular}
\end{minipage}
	\caption{Evolution of the average {(blue)} and standard deviation  {(orange)} of different metrics of the test dataset per episode of: the reward value (first row), the energy error (second row), and the computation time (third row) for the global training. The top five performing models are shown in the top row in red. A table is shown with the corresponding training and simulation parameters. }
	\label{fig:trainingglobal}
\end{figure}

We first train the RL method for 500 episodes and evaluate the performance of the models using the reward at each episode. In Figure \ref{fig:trainingglobal}, we present the results of this global training. The rows in the figure represent the reward value, energy error, and computation time for each episode. In blue we show the average value obtained from the test cases, and in orange the standard deviation. We also show the total training time in the top left corner. We mark in red the five episodes
with the largest reward and choose the best-performing model among
those.

\begin{figure}
	\begin{minipage}{\linewidth}
		\begin{center}
			\centerline{\includegraphics[width=\textwidth]{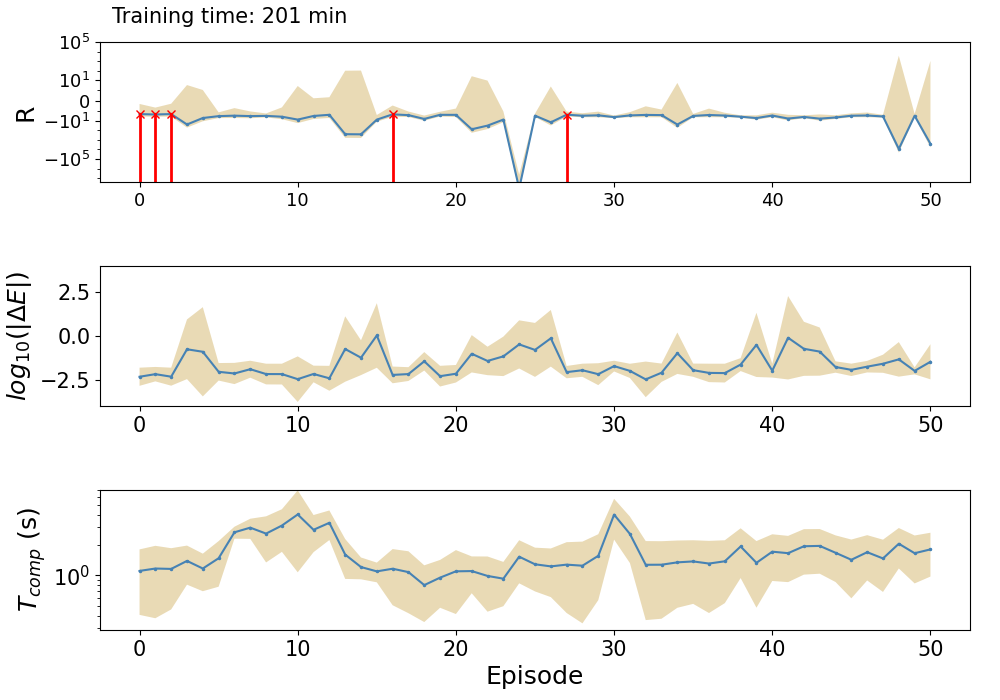}}
		\end{center}
	\end{minipage}
	\vspace{-20pt} 
	$$$$\hspace{20pt}
	\begin{minipage}{0.4\linewidth}
		\begin{tabular}{lclc}
			\hline
			\textbf{Local search 1}&\\
			\hline
			Number of stars & 5& \hspace{20pt}
			Max episodes & 50 \\
			Learning rate & $1\times 10^{-4}$& \hspace{20pt}
			Model chosen & 27\\
			\hline
		\end{tabular}
	\end{minipage}
	\caption{Evolution of the average  {(blue)} and standard deviation  {(orange)} of different metrics of the test dataset per episode of: the reward value (first row), the energy error (second row), and the computation time (third row) for the local training performed after the global one. The top five performing models are shown in the top row in red. A table is shown with the corresponding training and simulation parameters. }
	\label{fig:traininglocal1}
\end{figure}
\begin{figure}
	\begin{minipage}{\linewidth}
		\begin{center}
			\centerline{\includegraphics[width=\textwidth]{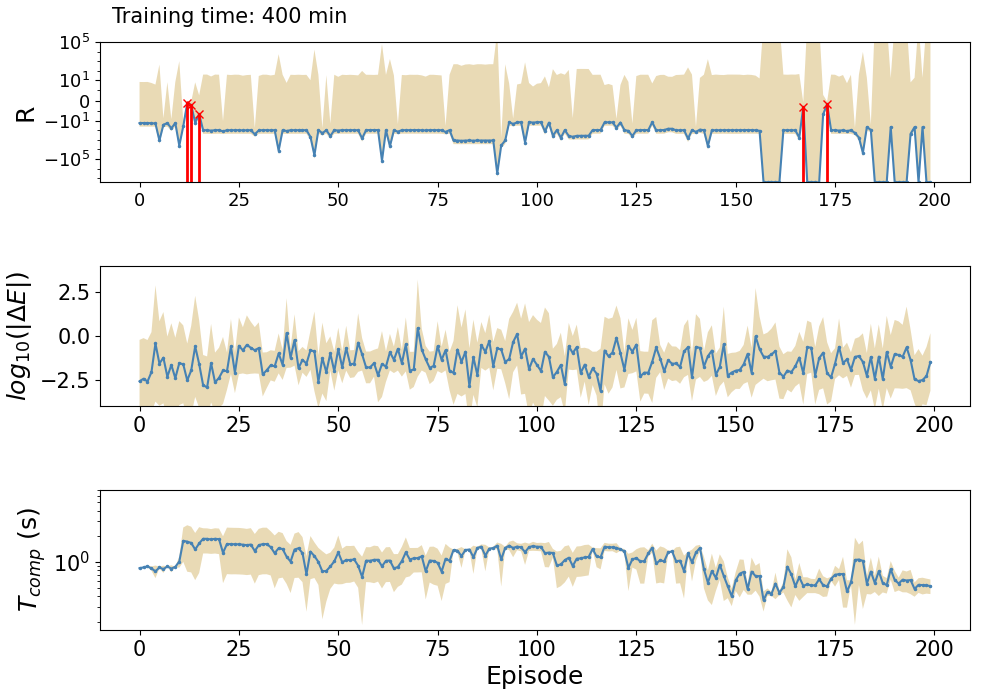}}
		\end{center}
	\end{minipage}
	\vspace{-20pt} 
	$$$$\hspace{20pt}
	\begin{minipage}{0.4\linewidth}
		\begin{tabular}{lclc}
			\hline
			\textbf{Local search 2}&&&\\
			\hline
			Number of stars & [5-20]& \hspace{20pt}
			Max episodes & 200 \\
			Learning rate & $1\times 10^{-4}$& \hspace{20pt}
			Model chosen & 173\\
			\hline
		\end{tabular}
	\end{minipage}
	\caption{Evolution of the average  {(blue)} and standard deviation  {(orange)} of different metrics of the test dataset per episode of: the reward value (first row), the energy error (second row), and the computation time (third row) for the local training for different bodies. The top five performing models are shown in the top row in red. A table is shown with the corresponding training and simulation parameters. }
	\label{fig:traininglocal2}
\end{figure}
\begin{figure*}
	\centering
	\includegraphics[width=1.5\columnwidth]{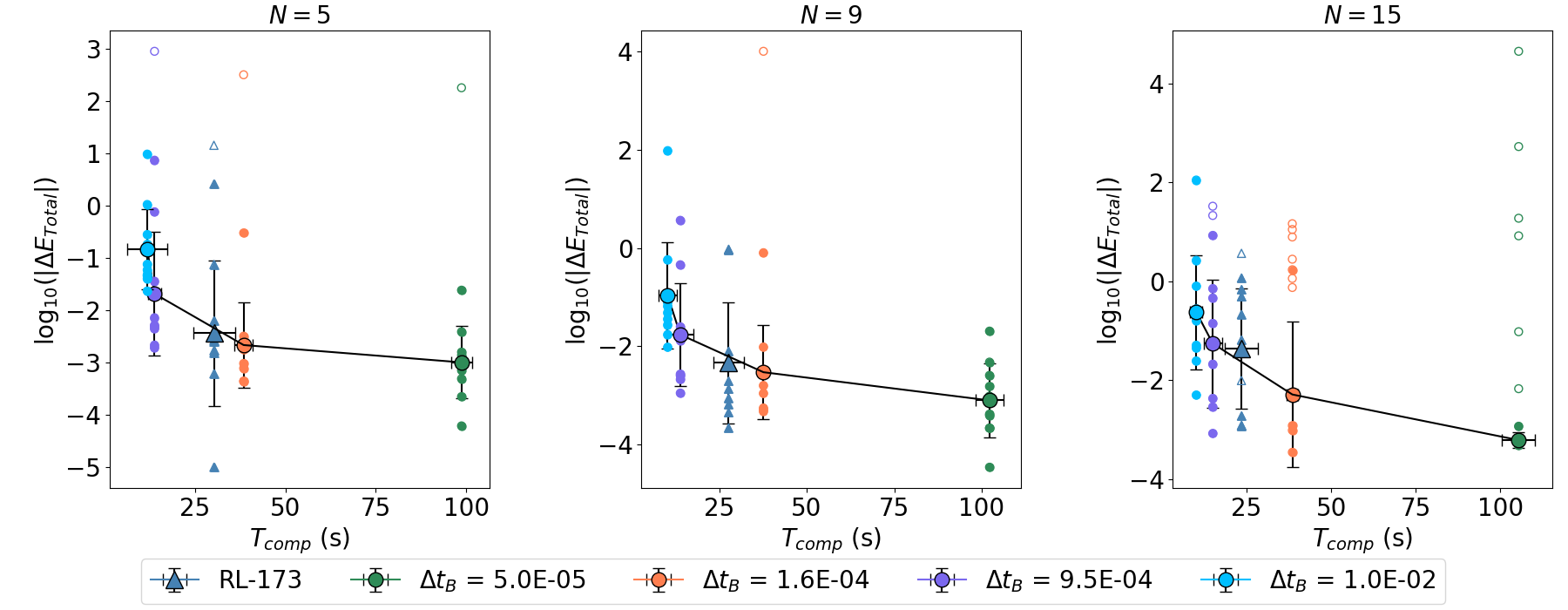}
	\caption{Average and standard deviation of the energy error and computation time for 10 different initializations run for 0.4 Myr. The results of the RL-173 model are compared to those of fixed $\Delta t_B$.}
	\label{fig:error2}
\end{figure*}

After the global training, we perform a local search (Figure
\ref{fig:traininglocal1}) starting from the best performing model
(model at episode 50 in Figure \ref{fig:trainingglobal}) for 50
episodes with a lower learning rate. Then, we choose the best-performing model; i.e., the one at episode 27. These trainings
are computationally inexpensive because the astronomical simulations
only contain five stars, but they are extremely useful to find a
pre-trained model to quickly achieve better performance. The next step consists of training ReLaTS on a different number of stars, to make the model more generalizable to a wider range of astrophysics problems. In Figure \ref{fig:traininglocal2}, we present the results of the training with a lower learning rate for 5 to 20 stars. We select the best-performing model, at episode 173, and compare the results in Figure \ref{fig:error2}.

\begin{figure}
	\centering
	\includegraphics[width=0.6\columnwidth]{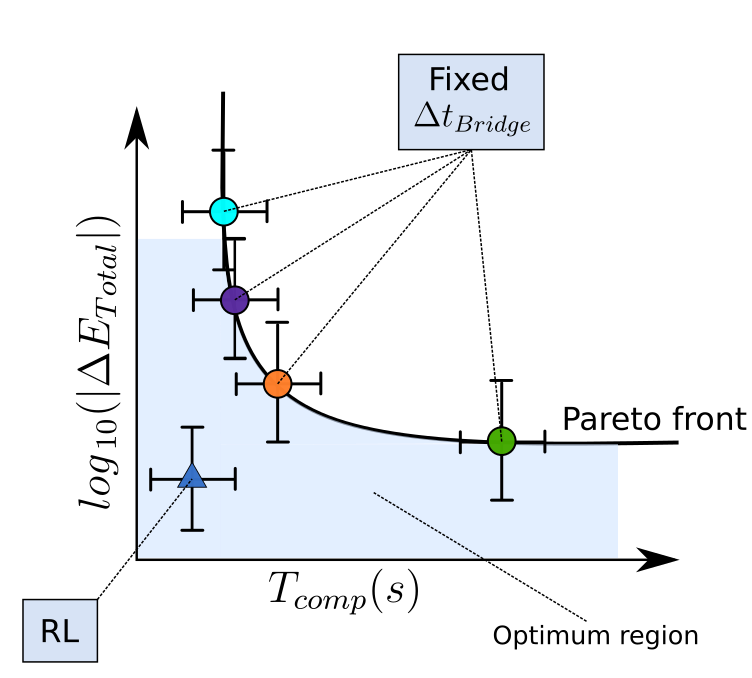}
	\caption{Schematic representation of the comparison of fixed $\Delta t_B$ with the RL method.}
	\label{fig:errorschem}
\end{figure}

To evaluate the performance of the selected model, we run multiple
simulations using the trained model and compare the results with those
without RL. Figure \ref{fig:errorschem} shows a schematic
representation of the plot that will be used for the statistical
comparison of the performance. The runs with different fixed $\Delta
t_B$ ideally form a Pareto front that ranges from the cases with large
computation time requirements and small energy errors (right of the
plot) compared to the ones with small computation times and large
errors (left of the plot). The Pareto front represents the best
performance that can be achieved with fixed time-step sizes. Results
below the curve represent better-performing cases compared to the
fixed $\Delta t_B$ cases. We aim to have a method that produces results on the Pareto front (effectively eliminating the expert knowledge) or below (improving the performance compared to a fixed time-step).

In Figure\,\ref{fig:error2}, we show the average and standard
deviation in computation time and energy error for 10 initializations
run for 0.4 Myr. We compare the results of the RL model at episode 173
(RL-173) to those with fixed $\Delta t_B$. We do that for 5, 9, and 15
stars. The energy errors obtained for each of the runs are shown as
points but for simplicity, the computation time is ignored
in the plot. An optimum value balances energy error (y-axis) and
computation time (x-axis). We show the Pareto front as a line joining
the mean value of the fixed time-step cases.

The unfilled markers in Figure\,\ref{fig:error2} represent those cases in which the planets escaped the planetary system.  As we discussed in section \ref{subsec:integration}, such dynamic reorganization is not currently supported by \texttt{Bridge}, and these simulations do not reflect a reliable measurement.

Model RL-173 performs approximately better in terms of energy conservation and computing time than the fixed time step simulations for $N = 5$ and $N=9$ compared to the fixed time-step case. The results become harder to interpret for $N = 15$, as for some cases with fixed $\Delta t_B$, most samples involve escaped planets (represented by unfilled markers).  Nevertheless, the algorithm achieves a comparable performance to the cases with fixed $\Delta t_B$ and results in fewer cases with escaped planets.

We demonstrate that ReLaTS performs at least equally well compared
to the best-performing constant \texttt{Bridge} timestep case. When initializing a simulation, it is common to use the default values for $\Delta t_B$ which generally leads to suboptimal results. ReLaTS achieves optimal performance in terms of computational time and accuracy without the need for expert knowledge or a convergence study.

\subsection{Integration results}
\label{subsec:integration}

To better understand the performance and extrapolation capabilities of
model RL-173, we perform multiple experiments.

\begin{figure*}
	\centering
	\subfloat[]{\includegraphics[width=0.5\textwidth]{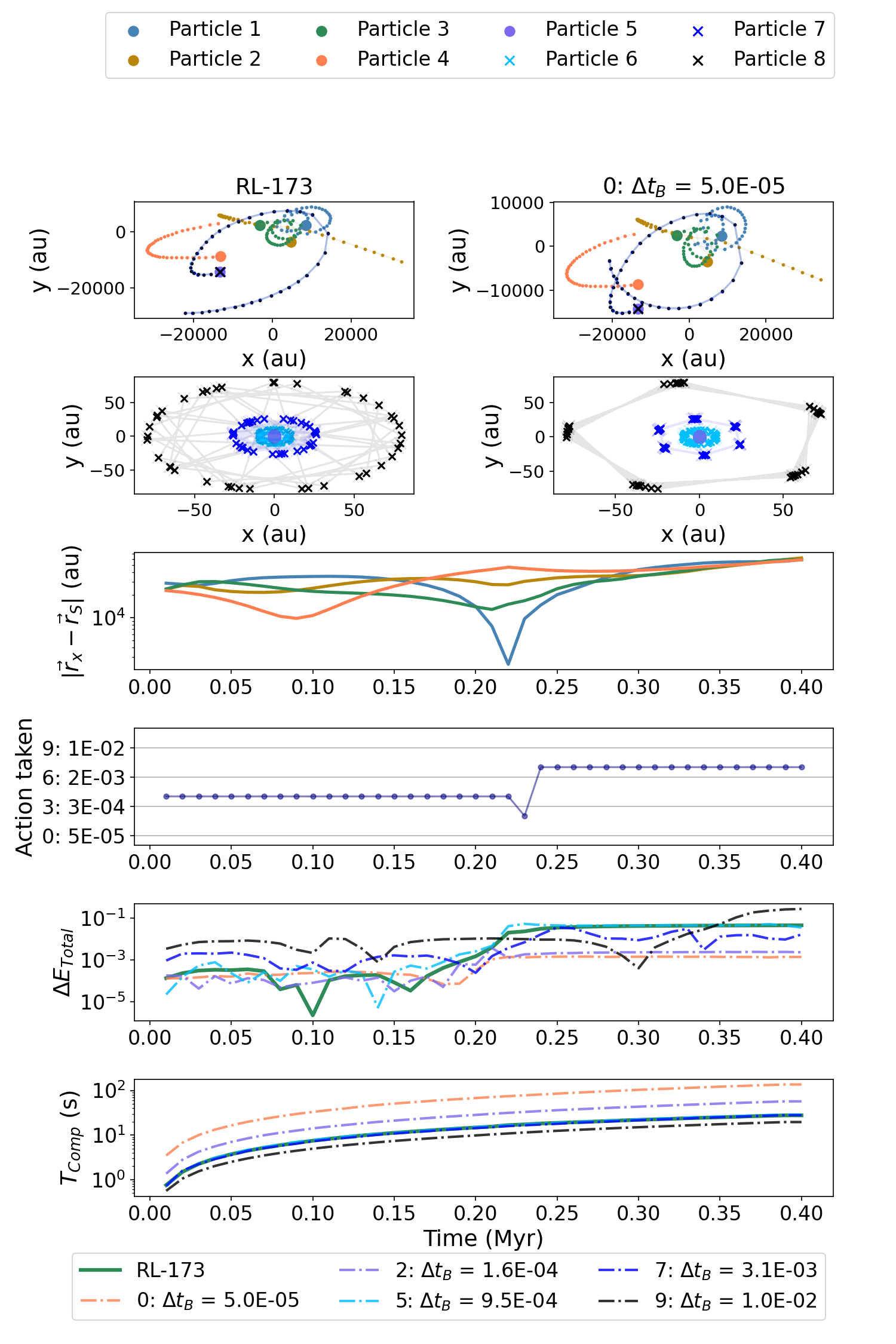}}
	\subfloat[]{\includegraphics[width=0.5\textwidth]{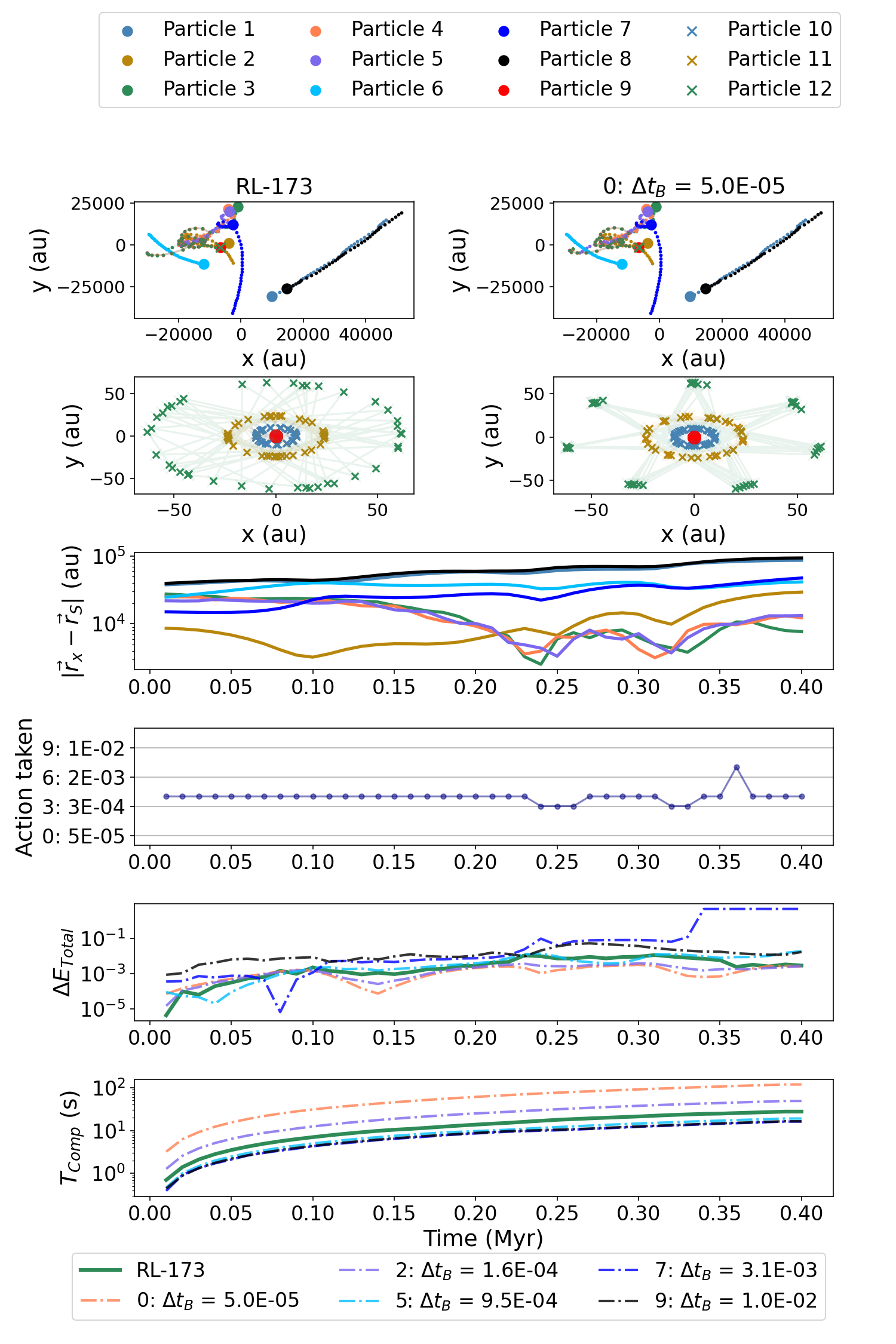}}
	\caption{Comparison of fixed $\Delta t_B$ to our RL model for 40 time steps (0.4 Myr). We present the trajectory in Cartesian coordinates of the star cluster (top-row panels) and the planetary system (second-row panels), the distance between each star to the one containing the planetary system (third row), the actions taken by the RL algorithm (fourth row), the energy error at each time step for each study case (fifth row), and the computation time for each study case (last row), for initializations with seed 4 (a) and seed 2 (b).}
	\label{fig:comparisonRLvsfixedsize}
\end{figure*}

We show the individual behavior of the model for initializations with
seed 4 (Figure \ref{fig:comparisonRLvsfixedsize} (a)) and seed 2
(Figure \ref{fig:comparisonRLvsfixedsize} (b)) with different numbers
of stars. The top row represents the position of the bodies in the
star cluster. The left column shows the evolution using the RL model
and the right one the results with the best-performing model with
fixed $\Delta t_B$. The second row presents the evolution of the
planetary system around its host star. The third row is the distance
from each star to the one with the planetary system. The fourth row is
the actions taken by the RL model at each step. The last two rows show
the energy error and computation time of each integration case (RL
model and several fixed $\Delta t_B$).

We see in Figure \ref{fig:comparisonRLvsfixedsize} (a) that there is only one close encounter, and that the RL model recognizes it and selects a more restrictive action (smaller time step). After the close encounter, the stars move further away from each other, and the model chooses a less restrictive action, saving computation time. In Figure \ref{fig:comparisonRLvsfixedsize} (b), we see a case for 9 stars. We recognize two or three close encounters and see that the RL model adapts accordingly. Finally, it achieves an energy error in the lowest range compared to the fixed $\Delta t_B$ cases without incurring large computation times. 

\begin{figure*}
	\centering
	\subfloat[]{\includegraphics[width=0.5\textwidth]{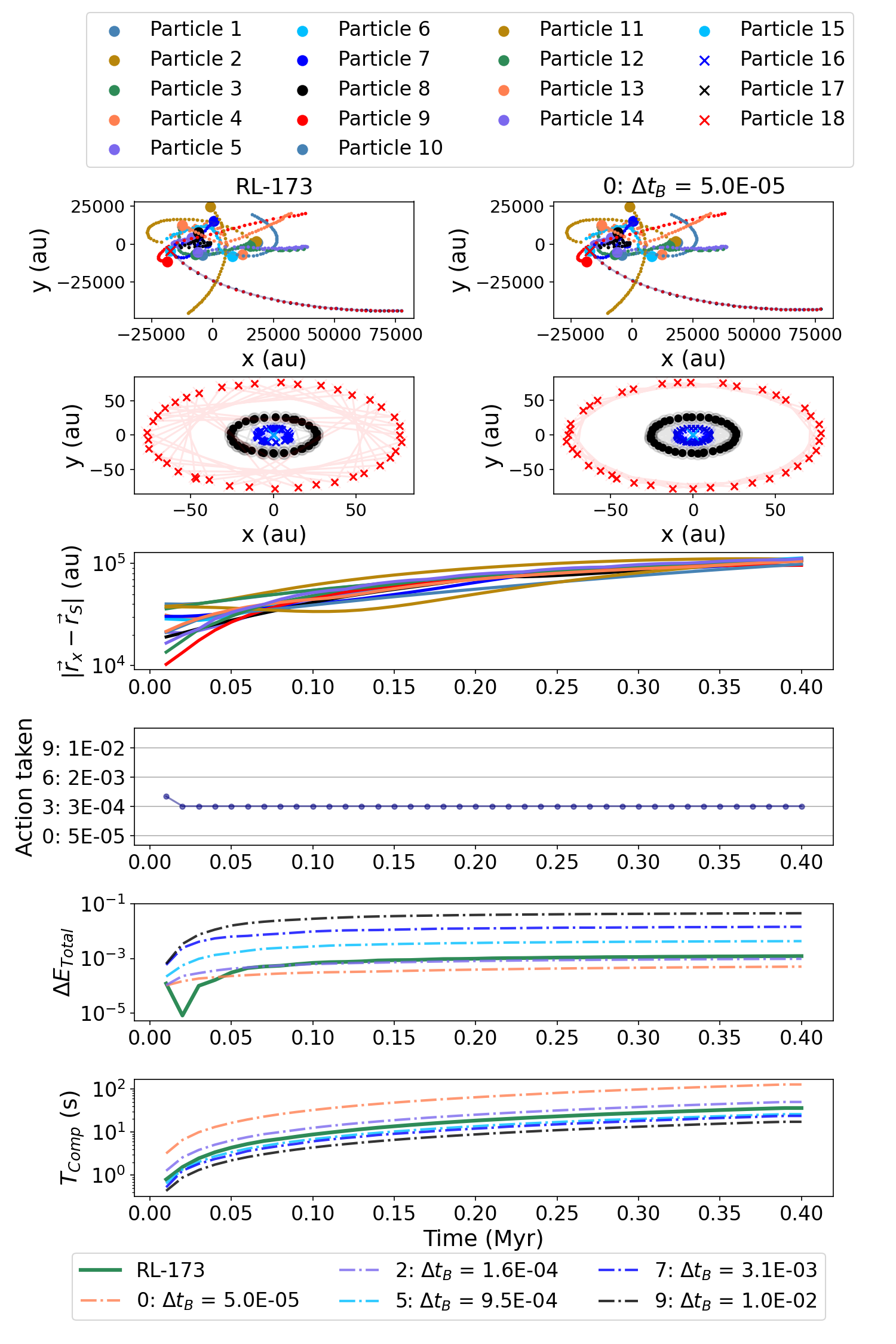}}
	\subfloat[]{\includegraphics[width=0.5\textwidth]{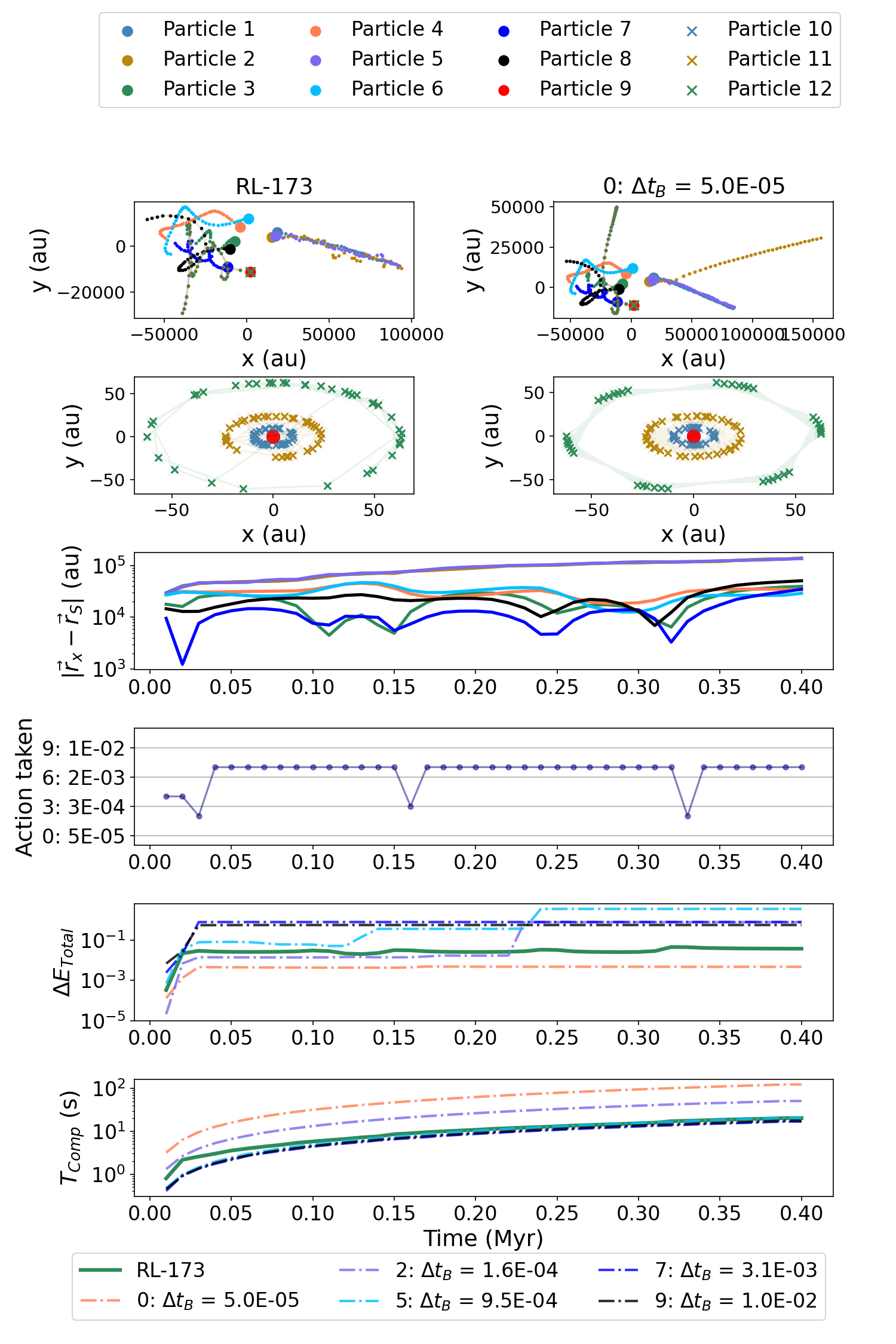}}
	\caption{Comparison of fixed $\Delta t_B$ to our RL model for 40 time steps (0.4 Myr). We present the trajectory in Cartesian coordinates of the star cluster (top-row panels) and the planetary system (second-row panels), the distance between each star to the one containing the planetary system (third row), the actions taken by the RL algorithm (fourth row), the energy error at each time step for each study case (fifth row), and the computation time for each study case (last row), for two initializations with seeds 3 (a) and 4 (b).}
	\label{fig:comparisonRLvsfixedsize2}
\end{figure*}

Figure \ref{fig:comparisonRLvsfixedsize2} (a), shows an example with
15 stars without any close encounters. In this case, the algorithm
learns to keep a constant action, balancing energy error and
computation time. In Figure \ref{fig:comparisonRLvsfixedsize2} (b),
the model recognizes close encounters and adapts accordingly by
selecting a higher value of the actions, obtaining an energy error
that is smaller than most of the other fixed $\Delta t_B$ cases with a
computation time comparable to the largest constant \texttt{Bridge} time step.

\begin{figure*}
	\centering
	\subfloat[]{\includegraphics[width=0.5\textwidth]{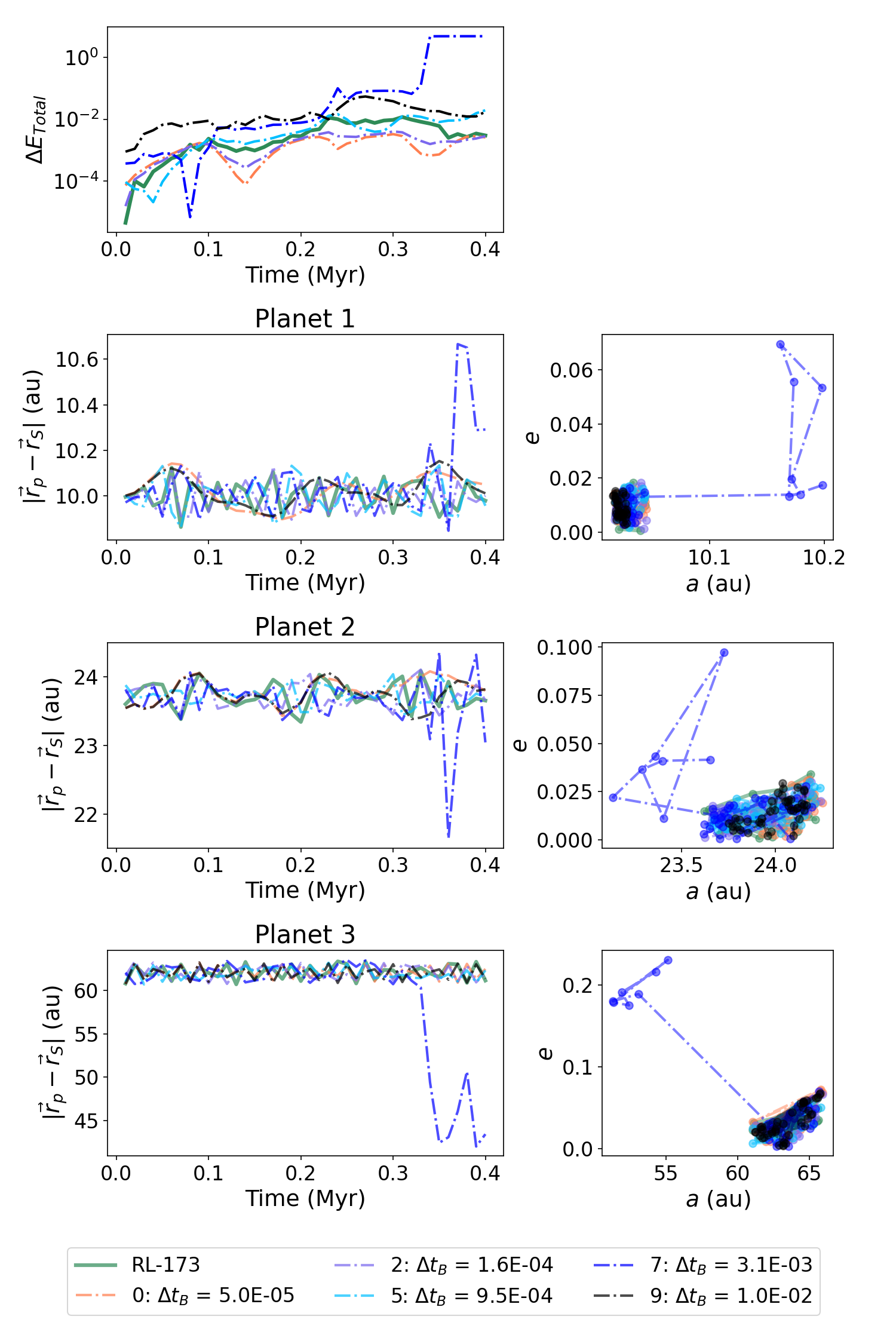}}
	\subfloat[]{\includegraphics[width=0.5\textwidth]{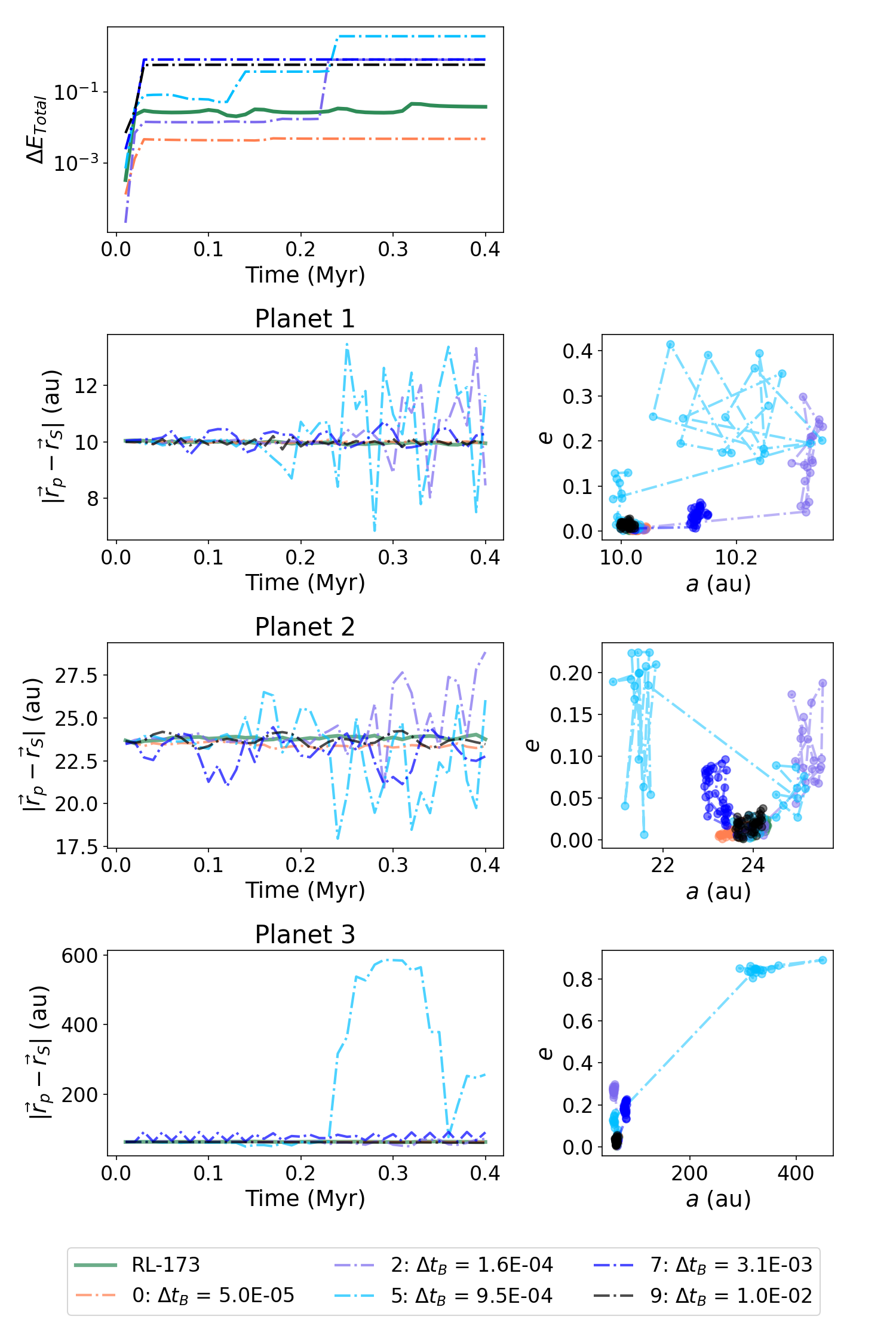}}
	\caption{Comparison of fixed $\Delta t_B$ to our RL model for 40 time steps (0.4 Myr). We present the energy error (top row), the time evolution of the distance of each planet to their central star (left panels), and the evolution of the semi-major axis ($a$) against the eccentricity ($e$) (right panels) for each planet with seeds 2 (a) and 4 (b).}
	\label{fig:avse1}
\end{figure*}

In Figures \ref{fig:comparisonRLvsfixedsize} and
\ref{fig:comparisonRLvsfixedsize2}, we observe sudden jumps in the energy error for certain cases. To understand these jumps, we plot in Figure \ref{fig:avse1} the evolution of the distance between each planet and its host. We do this for the same scenarios as in Figures
\ref{fig:comparisonRLvsfixedsize} (b) and \ref{fig:comparisonRLvsfixedsize2} (b). Additionally, we present the
semi-major axis and eccentricity to assess the dynamical evolution of the planetary system. We observe how the jumps in energy error correspond to the distance of a planet to the central star as it increases radically. Similarly, we observe that the planet's eccentricity grows. Such variations can be internal, but for the outermost planet (planet 3) the changes in its orbit are induced by the other stars in the cluster.

We show in Figure \ref{fig:avse1} that the jumps in energy error
correspond to the planets moving further away from their central
star. Reinforcement learning helps prevent planets from escaping their
host star during the simulation (unfilled symbols in figure \ref{fig:error2}), indicating that the planets' escape results from large energy error during close encounters. The reason why RL helps to mitigate this drawback of \texttt{Bridge} methods, is that it allows to reduce the time step time during close encounters. This situation could be further solved by further reducing the \texttt{Bridge} time step to adapt to the needs of the problem. This may however, lead to impractically small time-step sizes and large computation times. In our method, we set lower and upper limits for the values of the bridge time step (i.e., for the actions) to avoid extreme values of $\Delta t_B$.

\section{Generalization and robustness for long time integration}
\label{sec:experiments}

We have seen that the trained RL model manages to achieve results that
are better than those with fixed $\Delta t_B$. Those results were
obtained with conditions similar to those used to train the
model. Therefore, we want to understand its performance when applied
to different scenarios. To do that, we carry out experiments in which
we modify the final integration time, the integrators used, and the
\texttt{Bridge} time-step parameter.

\subsection{Number of bodies}

In Figure \ref{fig:error2}, we showed a comparison of the performance
of the RL model for 10 different initializations for three cases of
$N$. We use this plot as a baseline with which to compare the other
experiments.

\subsection{Long term integration}
The model is trained on simulations that were run for 0.4 Myr. We study the performance of the trained model on longer integration times to understand the possible use of our method for long-term simulations. 

\begin{figure*}
	\centering
	\subfloat[]{\includegraphics[width=0.5\textwidth]{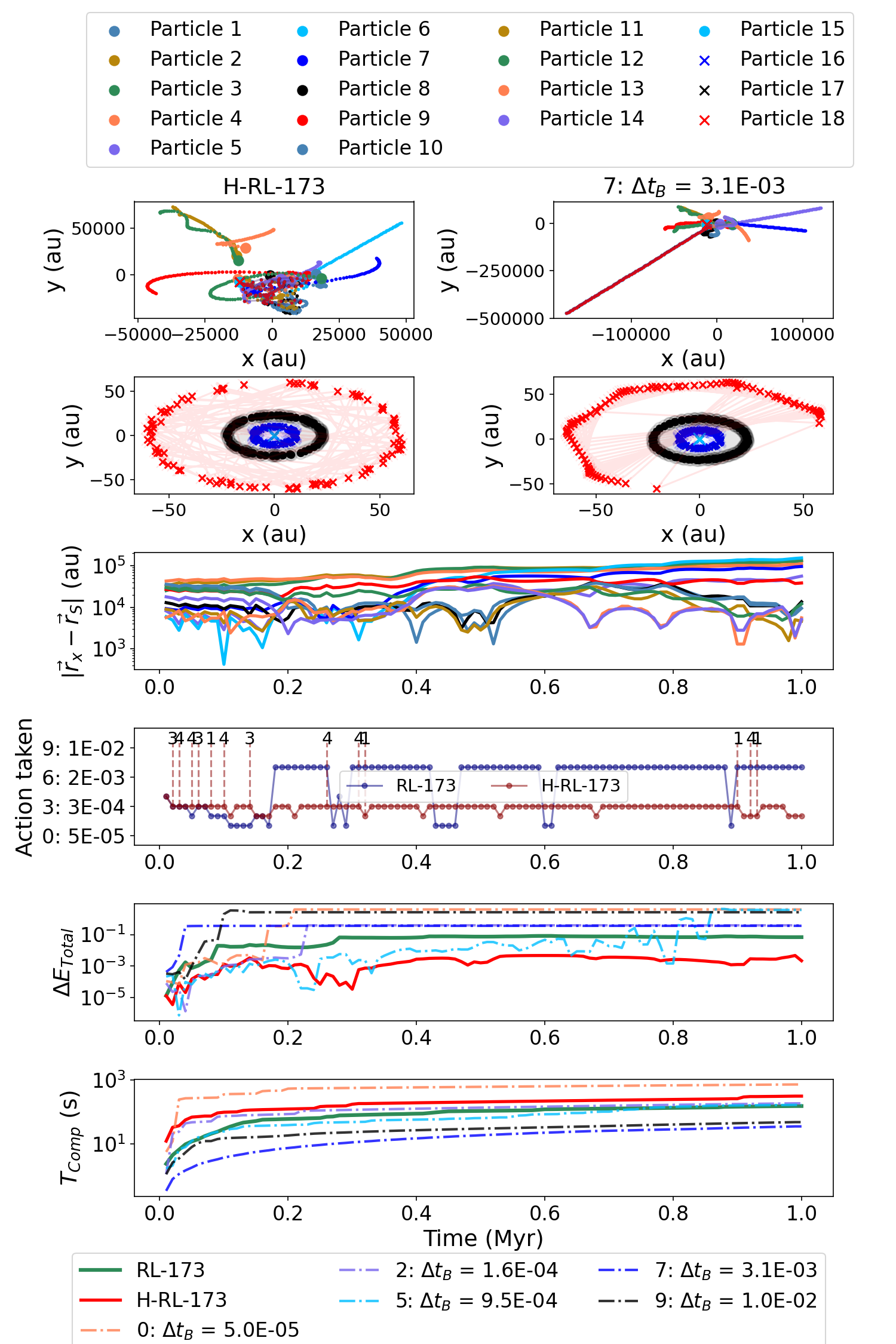}}
	\subfloat[]{\includegraphics[width=0.5\textwidth]{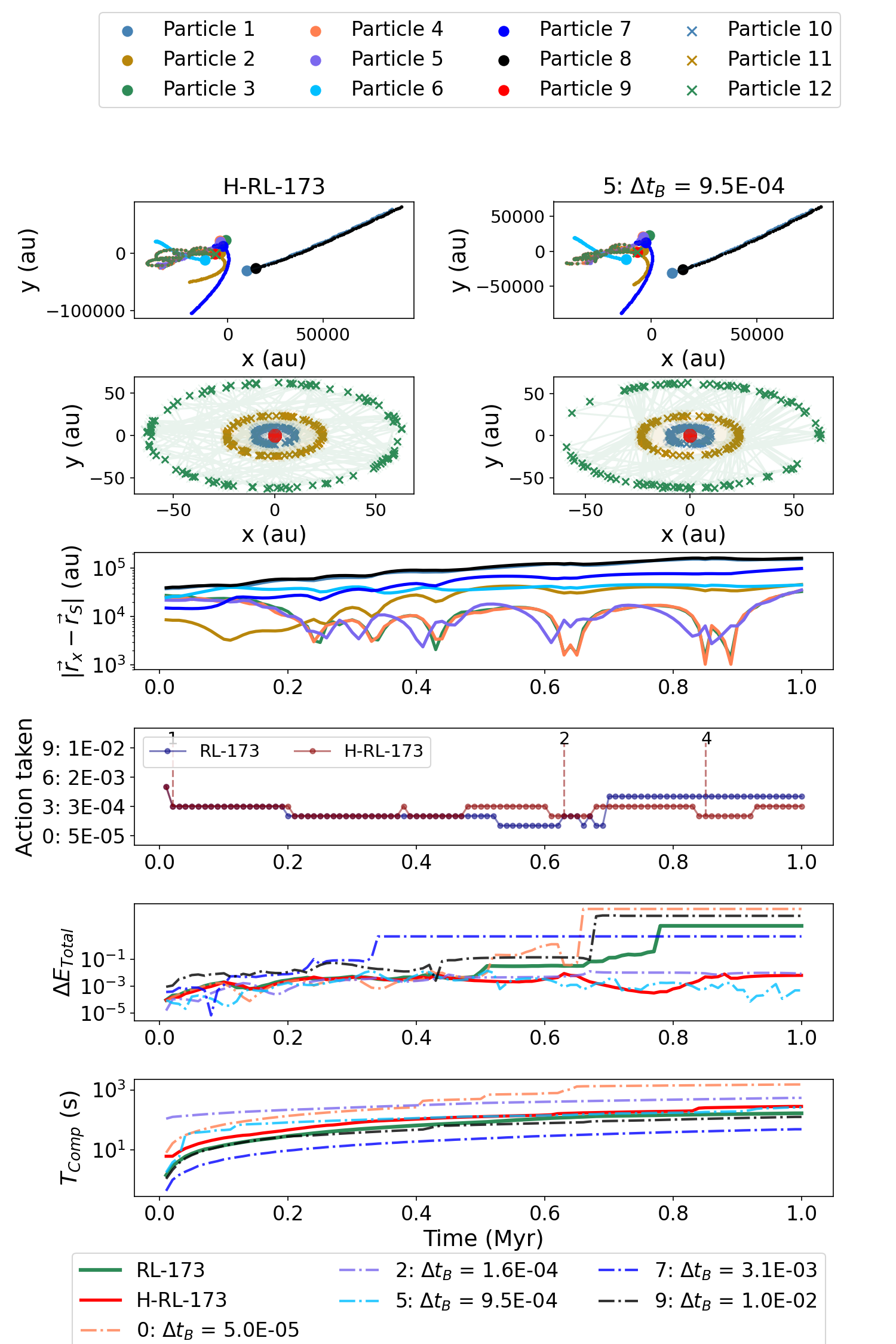}}
	\caption{Comparison of fixed $\Delta t_B$ to our RL and H-RL models for 100 time steps (1 Myr). We present the trajectory in Cartesian coordinates of the star cluster (top-row panels) and the planetary system (second row panels), the distance between each star to the one containing the planetary system (third row), the actions taken by the RL algorithm (fourth row), the energy error at each time step for each study case (fifth row), and the computation time for each study case (last row), for two initializations with Seeds 1 and 2.}
	\label{fig:comparisonRLvsfixedsizelong}
\end{figure*}

In Figure \ref{fig:comparisonRLvsfixedsizelong}, we show two examples
of the simulation with seeds 1 (a) and 2 (b) with different numbers of
bodies. The model identifies close encounters and
chooses more restrictive actions to keep the energy error
small. Figure \ref{fig:comparisonRLvsfixedsizelong} shows 
that the energy error is systematically smaller than with the
smaller but constant value of $\Delta t_B$ case while the computation cost
remains small. In Figure \ref{fig:comparisonRLvsfixedsizelong} (b), we
see a case without pronounced close encounters, but where some cases
experience a jump in energy error at $t\approx 0.7$ Myr. Making a
mistake in the choice of time-step size can lead to sudden changes in
energy error. The RL algorithm can keep the energy error small for
longer than other cases shown, but at $t\approx 0.8$ it still
experiences a jump.

For long-term integration, it is essential to avoid mistakes in the
choice of the time step. A wrong choice in the action by the RL model
can render a long simulation unusable. To prevent this, we propose a
method that identifies jumps in energy error and adapts the action
accordingly to correct for a wrong choice of time step. A similar idea
was introduced in \cite{ulibarrena2024hybrid}. We evaluate the energy
error with respect to the previous step. If this relative error is
larger than a predetermined value, we repeat the integration step with
a time-step size that corresponds to a smaller action or, if we were
already at the smaller action, reduce the time-step to half its
size. This process can be repeated until the energy error falls within
the desired limits. We choose a maximum of 4 iterations to avoid
incurring excessively large computation time.

The results obtained with this method, denominated as H-RL (for
Hybrid-RL) are included in Figure
\ref{fig:comparisonRLvsfixedsizelong}. The hybrid implementation
manages to prevent jumps in energy error. In the fourth row, we show
the actions taken by the H-RL method, and also at which steps the
hybrid method was activated
and the number of iterations it performed to lower the energy error
below the threshold.
This activation treshold is determined from the energy error in the integration:
\begin{equation}
\Delta t_B = \dfrac{\Delta t_B }{2} \qquad
\text{for} 
\qquad
\text{log}_{10}(\Delta E_i) - \text{log}_{10}(\Delta E_{i-1})> 0.3.
\label{eq:hybrid}
\end{equation}
The H-RL method results in energy errors orders
of magnitude smaller than the best result without incurring much
additional computation time (Figure
\ref{fig:comparisonRLvsfixedsizelong} (a)). For the cases where the RL
method experienced a jump in energy error
(\ref{fig:comparisonRLvsfixedsizelong} (b)), the hybrid method
prevents this jump, leading to a final energy error comparable to the
most accurate results for the constant time-step, and faster.

\begin{figure*}
	\centering
	\includegraphics[width=1.5\columnwidth]{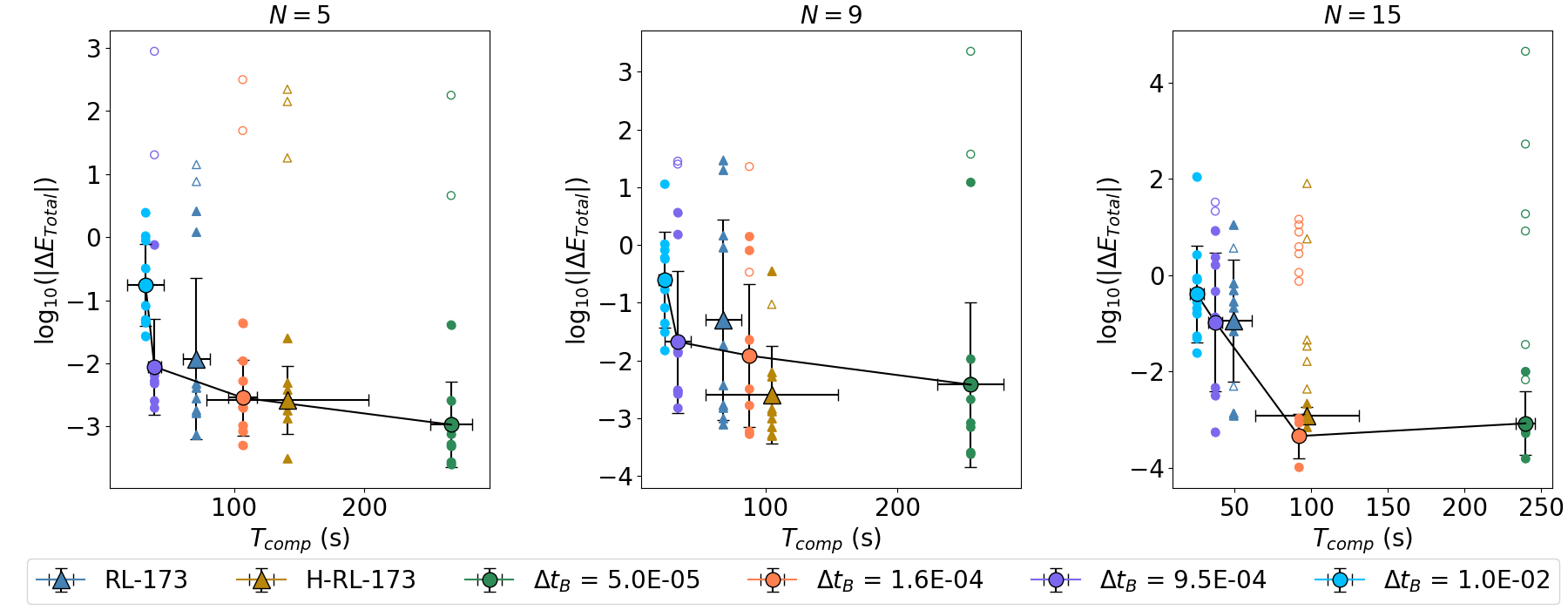}
	\caption{Average and standard deviation of the energy error and computation time for 10 different initializations run for 1 Myr. The results of the RL-173 and the H-RL-173 models are compared to those of fixed $\Delta t_B$. }
	\label{fig:error5}
\end{figure*}

In Figure \ref{fig:error5}, we present the statistical analysis for
the integration up to 1 Myr, including the RL and the H-RL results. We
find that for $N =5, 9,$ and $15$, the RL model results in a similar
performance to that of the fixed time-step cases. The mean value
appears over the Pareto front as the use of RL helps to prevent the
cases in which planets escape. In most cases of fixed $\Delta t_B$,
the values with a larger energy error have been discarded as they
involve escaping planets. This results in the mean energy error
displayed in the plots being smaller than for RL.  The H-RL case
further reduces the final energy error at the cost of some computation
time and leads to unequivocally better performance for $N = 9$. For $N
= 5$ and $15$, the results with H-RL are comparable to those with
fixed time-step size but with a larger standard deviation in
computation time. This represents an improvement in computation time
with respect to the fixed-step cases for similar values of the energy
error.

The hybrid method is particularly good for long-term integration as an
increase in the energy error is rarely reversible. For simplicity, we
will not include the hybrid integrator results in the following
experiments.

\subsection{Application of a time-step parameter}

\begin{figure*}
	\centering
	\includegraphics[width=1.5\columnwidth]{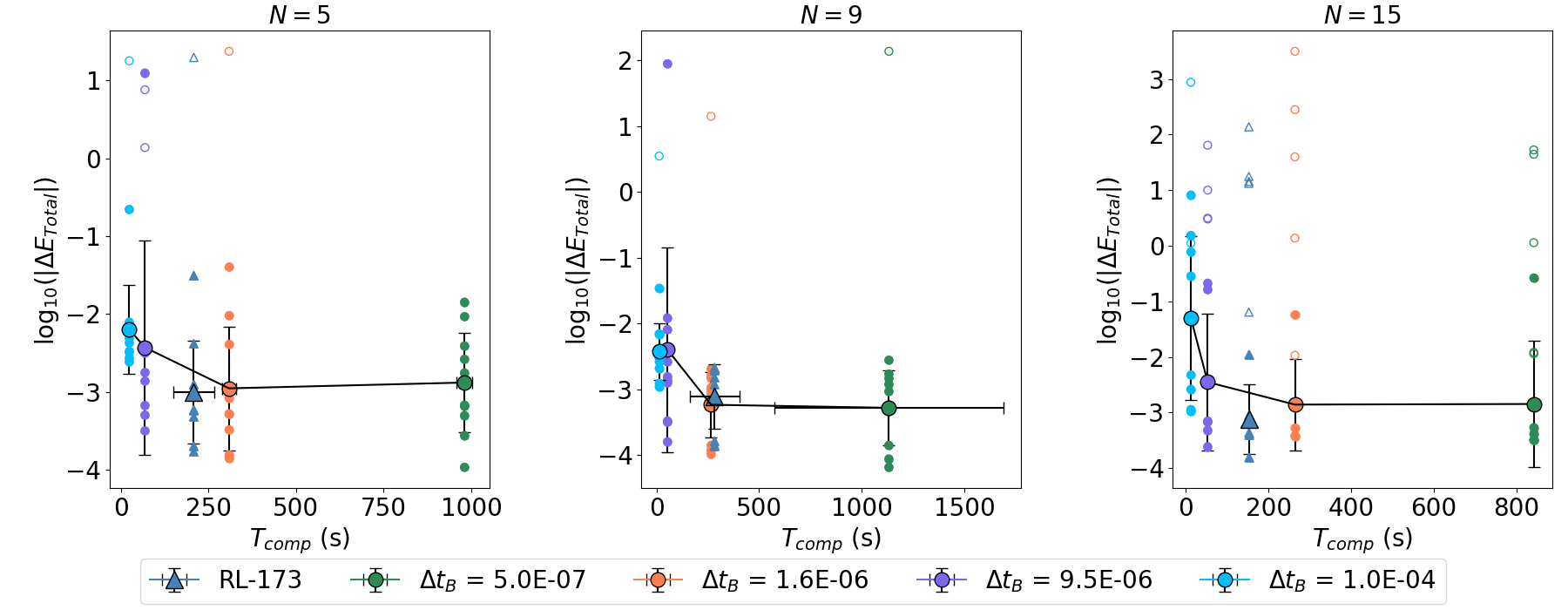}
	\caption{Average and standard deviation of the energy error and computation time for 10 different initializations run for 0.4 Myr. The results of the RL-173 model are compared to those of fixed $\Delta t_B$. The time-step parameter is changed to $10^{-2}$.}
	\label{fig:error4}
\end{figure*}

A time-step parameter ($\eta$) is used in integrators such as
\texttt{Hermite} and \texttt{Huayno} to scale the size of the time steps to tune the simulations. For large values of $\eta$, the are simulations are faster than for lower values, at the cost of accuracy. We implement a similar feature to scale the values of $\Delta t_B$. In Figure \ref{fig:error4}, we show that the performance does not decrease by scaling the actions by $10^{-2}$. For all cases, the RL method performs better, or similarly, to the fixed-size cases. We observe here that scaling $\Delta t_B$ also results in fewer escaping planets.

\section{Knowledge transfer capabilities of the network}

\subsection{Numerical integrators}

One of the main limitations of reinforcement learning methods is their
difficulty extrapolating to different setups. We, therefore, want to
understand whether our trained model is independent of the choice of
integrators for the parent and child. We replace the cluster
integrator with Hermite and the planetary system integrator
with Ph4.

\begin{figure*}
	\centering
	\includegraphics[width=1.5\columnwidth]{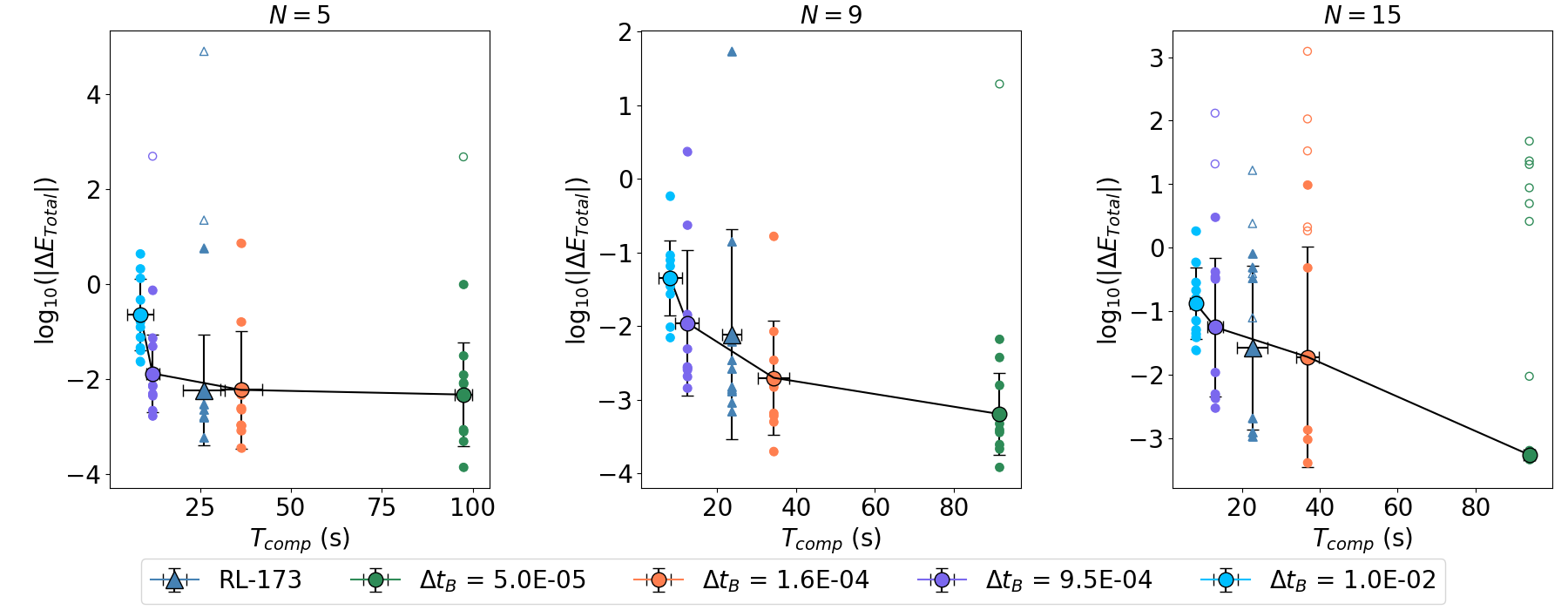}
	\caption{Average and standard deviation of the energy error and computation time for 10 different initializations run for 0.4 Myr. The results of the RL-173 model are compared to those of fixed $\Delta t_B$. The numerical integrators used in this case are different from those used for training. }
	\label{fig:error3}
\end{figure*}

In Figure \ref{fig:error3}, we show how the performance of the RL
model remains comparable to the baseline case by reaching better
energy conservation at greater speed for $N = 5, 15$. For $N = 9$, the
average for the RL method is located approximately on the Pareto
front, indicating that changing the integrator algorithm does not affect the reinforcement learning model's behavior. The network is able to generalize across numerical integrators, making ReLaTS independent of the numerical scheme.

\subsection{Star cluster simulation using Tree codes}\label{sec:tree}

We tested RL and H-RL methods on simulations using direct integrators
for the parent and the child and a small \textit{N}. To further test
the general applicability of the RL-\texttt{iBridge} methods, we use the Barnes-Hut Tree algorithm (\cite{barnes1986hierarchical}) to
integrate a cluster of $N = 1,000$ star in a Fractal distribution
(dimension 1.6) for 1 Myr (using initial seed 2).  The planetary
system orbiting one of the stars is still integrated using \texttt{Huayno}.

\begin{figure}
	\centering
	\includegraphics[width=\columnwidth]{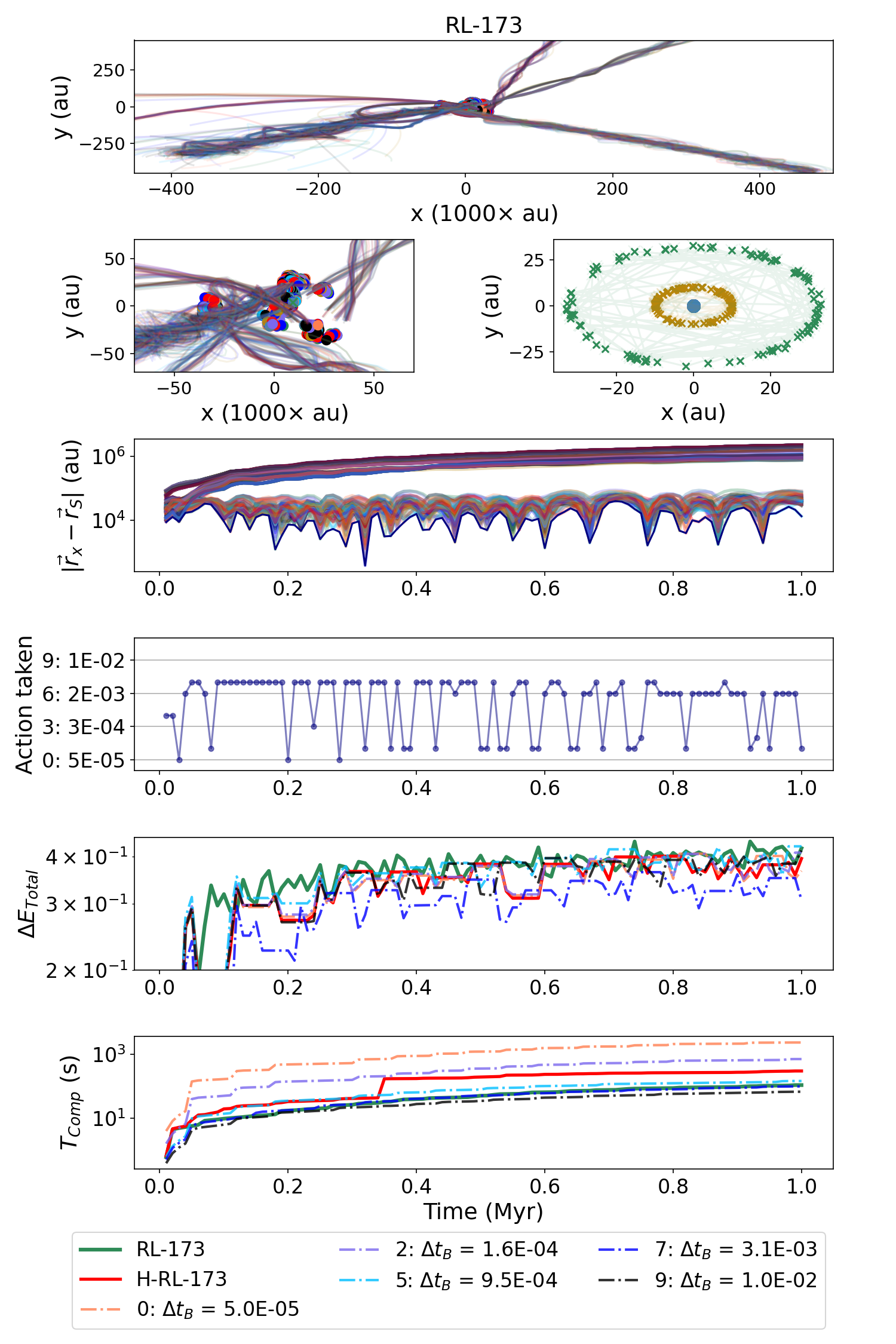}
	\caption{Comparison of fixed $\Delta t_B$ to our RL model for 100 time steps (1 Myr). We present the trajectory in Cartesian coordinates of the star cluster (top-row panel), a close-up view of the star cluster evolution (second row, left panel), and the trajectory of the planetary system (second-row, right panel). The distance between each star to the one containing the planetary system is shown in the third row, the actions taken by the RL algorithm in the fourth row, the energy error at each time step for each study case in the fifth row, and the computation time for each study case in the last row. The simulation is run for Seed 1 and BHTree integrator for the star cluster evolution.}
	\label{fig:treecode}
\end{figure}

In Figure \ref{fig:treecode} we demonstrate that RL-173 is capable of
identifying close encounters and adapting the actions accordingly. The
H-RL method further reduces the energy error at the cost of
computation time. Both methods achieve results that balance accuracy
and computation time without the need for expert knowledge.

\subsection{Application to different study cases: proto-planetary disk in a triple}
\label{sec:disk}

We demonstrated how our method can be used to simulate star
clusters that include (at least one) planetary system.  We take a
triple system of stars with masses $m_{1,2,3} = [1 , 0.5, 0.5]\;
M_Sun$. The positions and velocities of the stars are initialized
using a fractal cluster model (\cite{goodwin2004dynamical}) with
fractal dimension 1.6 and a virial ratio of 0.5. A proto-planetary
disk is placed around a randomly selected star.  The planetary system's parameters are listed in Table
\ref{table:initialconditionsprotodisk}.

\begin{table}
	\vskip 0.15in
	\begin{center}
		\footnotesize
		\begin{sc}
			\caption{Initial conditions and integration parameters for the star system and proto-planetary disk. }
			\label{table:initialconditionsprotodisk}
			\hspace{-20pt}
			\begin{tabular}{lc}
				\textbf{Triple star}&\\
				\hline
				Number of stars & 3\\
				Masses & [1, 0.5, 0.5] $M_{Sun}$\\
				Virial ratio & 0.5 \\
				Fractal dimension & 1.6\\
				\hline 
				&\\
				\textbf{Proto-planetary disk}&\\
				\hline
				Inner disk radius & 10 au\\
				Outer disk radius & 1000 au\\
				Disk mass & 0.01 $M_{Sun}$\\
				\hline
				&\\
				\textbf{Integration}&\\
				\hline
				Triple star code & Huayno\\
				Cluster code $\Delta t$ & $10\; yr$\\
				Proto-planetary disk code & Huayno\\
				Planetary system code $\eta_P$ & $3\; yr$\\
				Check step size & $100$ $yr$\\
				Action range & $[10, 100]\; yr$ \\
				\hline
			\end{tabular}
		\end{sc}	
	\end{center}
	\vskip -0.1in
\end{table}

\begin{figure*}
	\centering
	\subfloat[]{\includegraphics[width=0.5\textwidth]{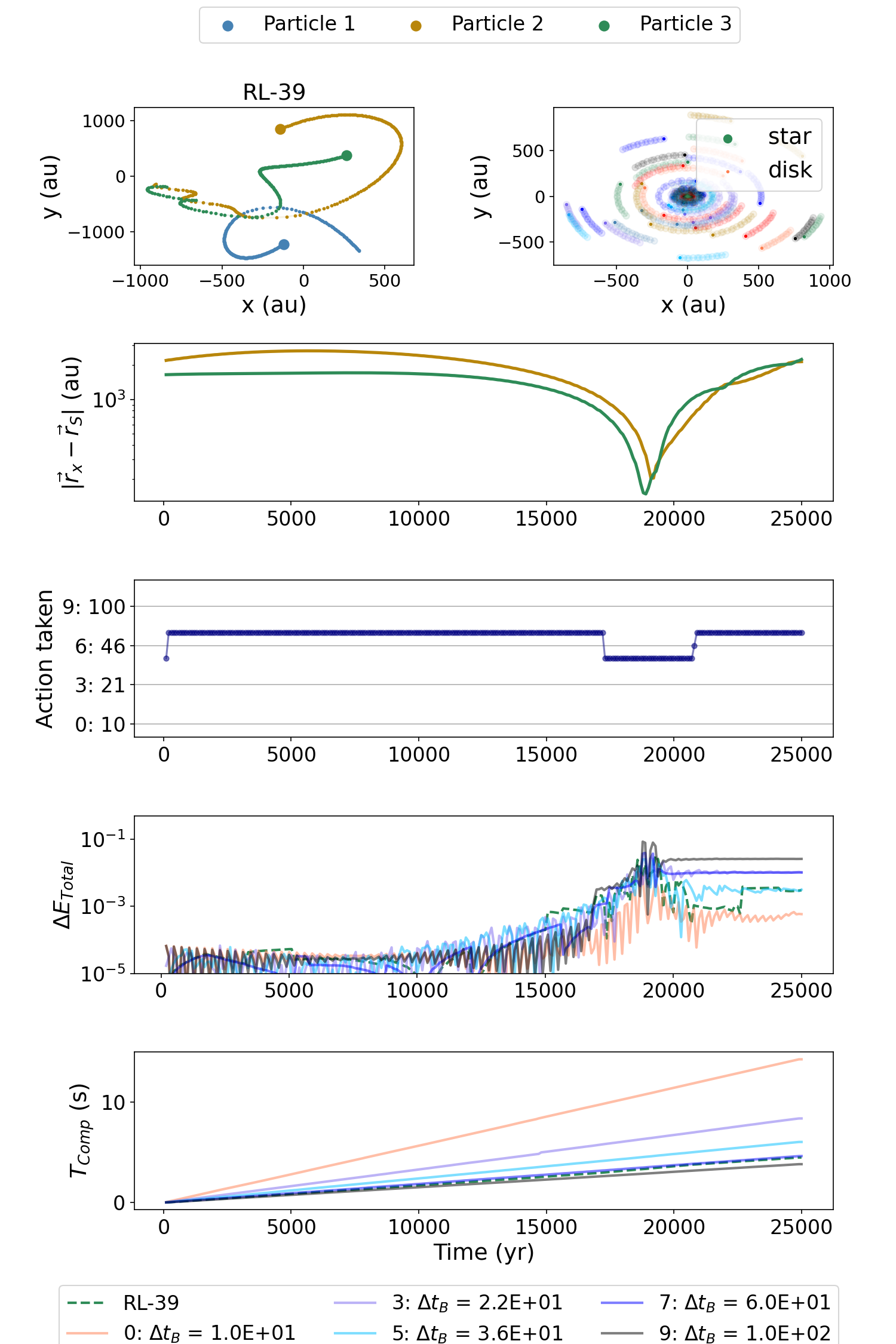}}
	\subfloat[]{\includegraphics[width=0.5\textwidth]{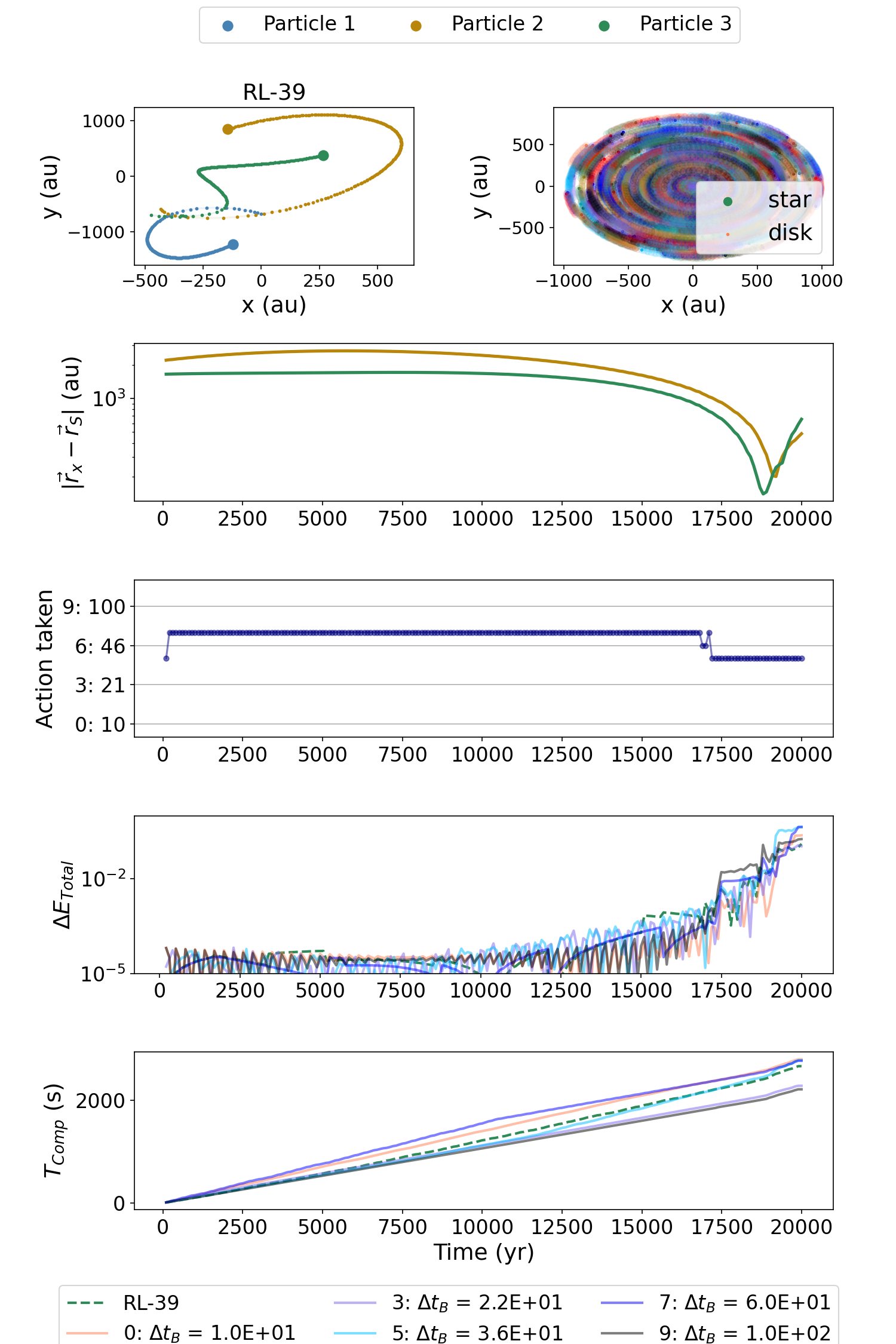}}
	\caption{Comparison of fixed $\Delta t_B$ to our RL model for 250 (a) and 200 (b) time steps (2500 and 2000 yr, respectively). We present the trajectory in Cartesian coordinates of the triple star (top-left panel) and the star and protoplanetary disk (top-right panel), the distance between each star to the one containing the planetary system (second row), the actions taken by the RL algorithm (third row), the energy error at each time step for each study case (fourth row), and the computation time for each study case (last row), for Seed 2345679. The disk contains 50 particles (a) and 2,000 particles (b).}
	\label{fig:protoplanetarydisk}
\end{figure*}

The scale on which this system evolves is different from that of
previous cases. We must therefore adapt the time-step sizes of the
integrators, the check step size, and the range of the actions. The
values used can be seen in Table
\ref{table:initialconditionsprotodisk}. Since the scale of the system
is completely different, we retrain the network for 50 episodes using
a learning rate of $10^{-4}$ and, to speed up the
training process the disk has only 50 particles. We show the results
of the use of the re-trained model (model 39) in Figure
\ref{fig:protoplanetarydisk}. For 50 particles (Figure
\ref{fig:protoplanetarydisk} (a)), the trained model identifies the
close encounter and adapts the actions to minimize the energy
error. The RL case achieves energy errors and computation times on the
lowest range. This model trained for 50 particles can be used for
larger systems. We therefore show in Figure \ref{fig:protoplanetarydisk}
(b) the results of applying this fast-trained model for a more
computationally-intensive process. We observe that the model still
yields better performances and conserved energy better than
simulations with a constant time step.

\section{Discussion}
\label{Sec:Discussion}

\begin{figure}
	\centering
	\includegraphics[width=\columnwidth]{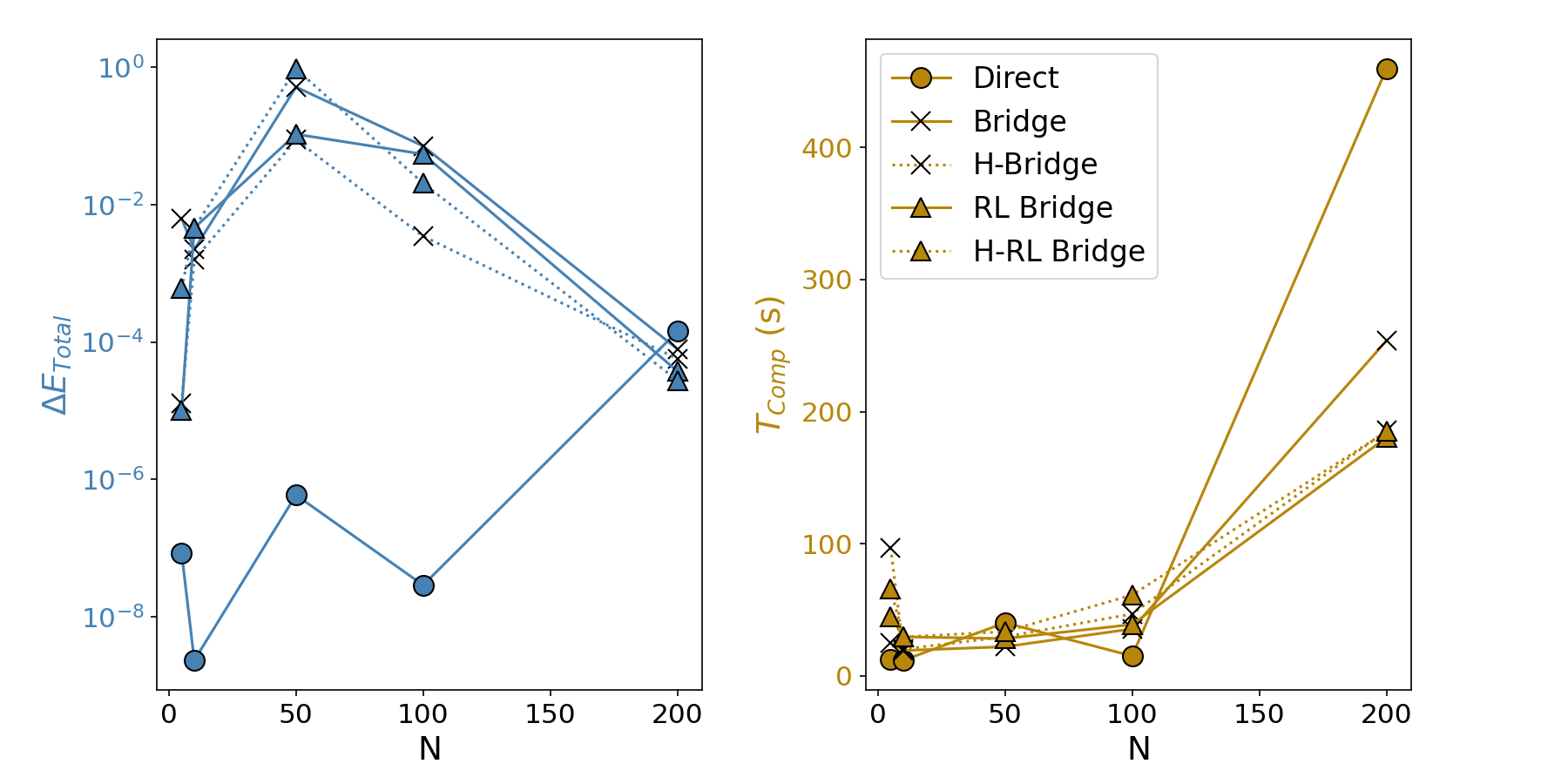}
	\caption{Comparison of the total energy error and computation time for an initialization with seed 3 run 40 steps with 9 stars. We compare the results with direct integration, with our \texttt{iBridge}, a hybrid implementation of the \texttt{iBridge}, and the cases with RL and H-RL. }
	\label{Fig:direct}
\end{figure}

This study examines star clusters containing 5 to 15 stars, one of
which hosts a planetary system with a variable number of planets. To compare the performance of the \texttt{Bridge} method with direct \textit{N}-body integration, Figure \ref{Fig:direct} presents the final energy error and computation time as functions of the cluster size $N$. Direct integration exhibits quadratic scaling in computational cost with the number of bodies, and linear scaling with the time step. For our experiment, the planetary system is integrated with a typical integration time step ($dt_{\rm pl}$) of about a month (for an inner semi-major axis of 10\,au as for the inner-most planet). For the star cluster, the typical dynamical time scale ($dt_{\rm cl}$) is on the order of 1\,Myr. In direct integration codes, the entire star cluster would have to be integrated with this small time step to ensure the accuracy of the integration of the planetary system, whereas much larger time steps could be adopted. This time-scale discrepancy is the main reason to adopt \texttt{Bridge}, in which we can separate the small (planetary) dynamics that is integrated on a time scale of months from the larger (cluster) dynamics which we integrate on a time scale of tens of thousands of years. Without \texttt{Bridge} the computational complexity of integrating the cluster with a $t=1$~Myr dynamical time scale is
\begin{equation}
  t_{\rm direct} \simeq \frac{t}{dt_{\rm pl}}(N_\star + N_{\rm planets}^2) 
\end{equation}
\begin{equation}
  t_{\rm Bridge} \simeq \frac{t}{dt_{\rm cl}} \times N_\star^2 + \frac{t}{dt_{\rm pl}} \times N_{\rm planet}^2.
\end{equation}
For example, for $N_\star = N_{\rm planets} = 5$ we then find $t_{\rm
  direct}/t_{\rm Bridge} \simeq 4$.  But for $N_\star =100$ (with the
same number of planets), the gain in speedup with \texttt{Bridge} has already increased to a factor 400.

Using \texttt{Bridge} has therefore enormous advantages in terms of
computational efficiency. However, the free parameter in \texttt{Bridge}, the so-called \texttt{Bridge} time step, $\Delta t_{B}$, requires expert knowledge to be selected for optimum integration speed and accuracy.  A too large value of $\Delta t_{B}$ will make the calculation inaccurate, whereas a too small value will make the simulation slow. Finding this optimum value for $\Delta t_{B}$ is often hard, and it may vary with time as the cluster or the planetary system evolves.

To take away the expert knowledge from the initialization of the system, we developed ReLaTS; the reinforcement learning algorithm to determine the cross integration (\texttt{Bridge}) time step $\Delta t_{B}$.

For comparison and tuning (see Figure \ref{Fig:direct}), we tested three different integrator configurations. Firstly, we performed the simulations by directly integrating all equations of motion in the same integrator (direct integration). Secondly, we applied a constant \texttt{Bridge} time step $\Delta t_B$; which is identical to the classic \texttt{Bridge} strategy as proposed in \cite{fujii2007bridge}.  Finally, we tested our variation of this method \texttt{iBridge}. In our reinforcement learning application, we explore two versions of the inclusive \texttt{Bridge} ({\tt iBridge}); one in which $\Delta t_B$ is determined at run time, and one in which, in addition, we allow the last step to be recomputed to correct a wrong choice made by the RL algorithm. This latter aspect can probably be improved by further training the model, and could be mitigated in future work by more robust training. 

Despite the small number of stars in our experiments, all
reinforcement learning \texttt{Bridge} calculations outperform the classic \texttt{Bridge} and (non-bridged) direct integration. In addition, the energy error is comparable or even better than in the classic cases (\texttt{Bridge} with expert-chosen constant time step).

Our training and test calculations, however, covered a rather limited
range of parameters. But considering the large dimensionality of
the problem, and the large number of free parameters, it would be very
elaborate to perform a systematic parameter search. We therefore
present ReLaTS, which we train on a limited set of problems while still being able to apply it to a larger variety of setups. 

ReLaTS is trained to optimally select the \texttt{Bridge} time step $\Delta t_B$ in such a way that computing time is optimal at minimal energy error.
However, energy error is not necessarily the optimal diagnostic. For example, when integrating a planetesimal disk orbiting a star, the integrator may make large errors in calculating the orbits of individual planetesimals without an appreciable variation of the energy of the entire system; simply because planetesimals are generally considerably less massive than the stars in
the cluster. 
Changing its orbit, or even if it collides with a planet, will hardly affect the total system energy. In future implementations, we consider
improving ReLaTS to include more robust integration quality criteria.

Testing any numerical integration of a Newtonian dynamical system is
hard, not only because of the previously -mentioned limitation of energy conservation as a diagnostic, but also because the system is chaotic. Integrating a system of multiple stars and planets to convergence (until we reach the correct solution) is extremely computationally time-intensive \citep{2015ComAC...2....2B}, and even then, any variation introduced by the numerical implementation drives an exponential divergence between the converged and the unphysical solution.  


Several assumptions have been made in this work. Energy error serves as the primary accuracy metric, indicating adherence to conservation laws but not guaranteeing fidelity in individual trajectories. Convergence is assessed via energy conservation, as dynamical convergence proves infeasible in chaotic systems. But energy error loses reliability as a diagnostic with increasing $N$ or for test-particle integrations.

The performance of ReLaTS matches or exceeds that of constant $\Delta t_B$ integrations. Distinct initializations produce substantial performance scatter under identical settings, reflecting sensitivity to time-step choices in chaotic dynamics. We have performed statistical analyses across realizations as no established implementation exists for dynamic $\Delta t_B$ adaptation in \texttt{Bridge} schemes; fixed small $\Delta t_B$ provides a baseline, though not invariably optimal. ReLaTS currently achieves the lowest energy errors at a satisfactory speed.

The \texttt{Bridge} approximation assumes separable subsystems. Figures \ref{fig:error2}, \ref{fig:error4}, and \ref{fig:error3} reveal occasional large energy errors across $\Delta t_B$ choices,
attributable to planetary escapes into the parent cluster. Such events
necessitate joint integration of interacting bodies. Advanced schemes
like \texttt{Nemesis} \citep{zwart2020non} allow multiple bridges to
cooperate, dynamical generation and destruction of new bridges, and
the exchange of objects across dynamic subsystems.  ReLaTS will prove
an essential asset for such a flexible dynamical reassessment at
runtime.

We further applied the reinforcement learning \texttt{Bridge} time step determination in ReLaTS to different integrators, and from 5 to 1000 stars in the cluster. We could have varied these parameters more
extensively, but we hope to be using ReLaTS in actual simulation
environments, where such testing will form a natural validation and
verification of the method as part of ongoing research. We further
experimented with using the network without retraining on a
hydrodynamical disk, but these experiments require further study and have therefore not been included in this work. 

The generality of ReLaTS permits application to diverse systems, such
as central stars with protoplanetary disks or gas clouds, without major modifications. The number of bodies may vary post-training, though performance may decline with increasing divergence from the training configuration.

\section{Conclusions}
\label{Sec:Conclusion}

We introduce ReLaTS, a reinforcement learning (RL)-based method that
automatically tunes the coupling time step in multi-scale simulations
of self-gravitating systems. Traditionally, these simulations use two
coupled integration codes; each handling different spatial and
temporal scales—connected by a ``{\em \texttt{Bridge} scheme}'' that
compromises between accuracy and compute time. The \texttt{Bridge} scheme is controlled by a constant time step, and selecting it requires expert knowledge.

ReLaTS replaces manual tuning by training an RL model to dynamically
select the optimal coupling time step ($\Delta t_{B}$) during
simulations. Tested on planetary systems embedded in stellar clusters
(spanning over three orders of magnitude in spatial and temporal
scales), the algorithm outperforms fixed time-step methods in energy
conservation and runtime efficiency.  The method adapts the time step
according to changing conditions, maintaining stable energy errors
over long integrations.

ReLaTS generalizes without retraining to varied astrophysical configurations, such as systems with different numbers of stars, planets, or even protoplanetary disks. The approach is numerical scheme-independent, meaning that it works with direct, symplectic, or Tree codes.  In this, ReLaTS provides a robust, adaptive, and expert-free solution to couple multi-scale gravitational simulations, improving both accuracy and computational efficiency while remaining broadly applicable.

\section{Acknowledgments}

It is a pleasure to thank Maxwell X. Cai for the insightful
discussions that led to this research.  This publication is funded by
the Dutch Research Council (NWO) with project number
OCENW.GROOT.2019.044 of the research programme NWO XL. It is part of
the project ``Unraveling Neural Networks with Structure-Preserving
Computing". Part of this publication is funded by the Nederlandse
Onderzoekschool Voor Astronomie (NOVA).

\section{Data Availability}

The code is publicly available at

\href{https://github.com/veronicasaz/RL_bridgedCluster}{github.com/veronicasaz/RL\_bridgedCluster}.

\section{Conflict of interest}
None


\end{document}